\documentclass[bibyear]{aa} 
\usepackage{amsmath}
\usepackage{graphicx}
\usepackage{media9}
\usepackage{xcolor}
\usepackage{natbib}
\usepackage{dirtytalk}
\usepackage{hyperref}
\usepackage{array,tabularx}
\usepackage{multimedia}
\usepackage[switch, mathlines]{lineno}
\usepackage{subfiles}
\usepackage{mathtools}
\usepackage{comment}
\DeclarePairedDelimiter\abs{\lvert}{\rvert}

\definecolor{myblue}{RGB}{51,90,140}
\newcommand{\myblue}{\textcolor{myblue}}

\usepackage[normalem]{ulem}

\begin{document}
\let\oldpageref\pageref
\renewcommand{\pageref}{\oldpageref*}

\title{\Large \textbf{Particle acceleration, escape and non-thermal emission from core-collapse supernovae inside non-identical wind-blown bubbles}}

\author{\textbf{\myblue{Samata Das}}\inst{1,2}\thanks{\myblue{samata.das@desy.de}}
\and{\textbf{\myblue{Robert Brose}}}\inst{3}
\and{\textbf{\myblue{Martin Pohl}}}\inst{1,2}
\and{\textbf{\myblue{Dominique M.-A. Meyer}}}\inst{2}
\and{\textbf{\myblue{Iurii Sushch}}}\inst{4,5}}

\institute{{Deutsches Elektronen-Synchrotron DESY, Platanenallee 6, 15738 Zeuthen, Germany} \and
{Institute of Physics and Astronomy, University of Potsdam, 14476 Potsdam, Germany} \and
{Dublin Institute for Advanced Studies, 31 Fitzwilliam Place, Dublin 2, Ireland} \and
{Centre for Space Research, North-West University, 2520 Potchefstroom, South Africa} \and
{Astronomical Observatory of Ivan Franko National University of L’viv, vul. Kyryla i Methodia, 8, L’viv 79005, Ukraine}}

\date{Received/
Accepted }

\abstract{
\noindent{\textbf{\myblue{Context.}} In the core-collapse scenario, the supernova remnants (SNRs) evolve inside the complex wind-blown bubbles, structured by massive progenitors during their lifetime. Therefore, particle acceleration and the emissions from these SNRs can carry the fingerprints of the evolutionary sequences of the progenitor stars.}\newline
{\textbf{\myblue{Aims.}} We time-dependently investigate the impact of the ambient environment of core-collapse SNRs
on particle spectra and the emissions, for two progenitors with different evolutionary tracks, accounting for the spatial transport of cosmic rays (CRs) and the magnetic turbulence which scatters CRs.}\newline
{\textbf{\myblue{Methods.}} We use the \textit{RATPaC} code to model the particle acceleration at the SNRs with progenitors having zero-age main sequence (ZAMS) masses of $20\,M_{\sun}$ and $60\,M_{\sun}$. We have constructed the pre-supernova circumstellar medium (CSM) by solving the hydrodynamic equations for the lifetime of the progenitor stars. Then, the transport equation for cosmic rays, and magnetic turbulence in test-particle approximation along with the induction equation for the evolution of large-scale magnetic field have been solved simultaneously with the hydrodynamic equations for the expansion of SNRs inside the pre-supernova CSM in 1-D spherical symmetry.\\
{\textbf{\myblue{Results.}} The profiles of gas density and temperature of the wind bubbles along with the magnetic field and the scattering turbulence regulate the spectra of accelerated particles for both SNRs. For the $60\,M_{\sun}$ progenitor the spectral index reaches 2.4 even below $10\,\mathrm{GeV}$ during the propagation of the SNR shock inside the hot shocked wind. In contrast, we have not observed  persistent soft spectra at earlier evolutionary stages of the SNR with $20\,M_{\sun}$ progenitor, for which the spectral index becomes 2.2 only for a brief period during the interaction of SNR shock with the dense shell of red supergiant (RSG) wind material}. At later stages of evolution, the spectra become soft above $\sim 10\,\mathrm{GeV}$ for both SNRs, as weak driving of turbulence permits the escape of high-energy particles from the remnants. The emission morphology of the SNRs strongly depends on the type of progenitors. For instance, the radio morphology of the SNR with $20\,M_{\sun}$ progenitor is centre-filled at early stages whereas that for the more massive progenitor} is shell-like.
}
    
\titlerunning{Particle acceleration in core-collapse SNR}
\authorrunning{S. Das et al.}
\maketitle

\section{Introduction}

Massive stars ($M_{\star} >\,8M_{\sun}$) shape the morphology of their circumstellar medium (CSM) through ionizing radiation and stellar winds \myblue{\citep{2003ApJ...594..888F, 2014ASTRP...1...61M,2022Galax..10...37D}}. As massive stars evolve through different stages, mass-loss in the form of stellar winds should vary \myblue{\citep{1987A&A...182..243M, 2012ARA&A..50..107L, 2015A&A...575A..60M, 2022arXiv220704786G}}, which is crucial for sculpting their environment \myblue{\citep{2005ApJ...630..892D, 2007ApJ...667..226D}}. The impact of stellar wind properties on the  structure of wind bubbles has been explored by many studies. For instance, \myblue{\citet{1996A&A...316..133G, 1996A&A...305..229G}} investigated the structure of wind bubbles around $35\,M_{\sun}$ and $60\,M_{\sun}$ stars in terms of the interacting stellar winds from different stages of evolution. The bipolar morphology of nebulae around Luminous Blue Variable (LBV) and Wolf-Rayet (WR) stars as a consequence of asymmetric ambient medium from prior evolutionary stages were studied in \myblue{\citet{Dwarkadas_2002, 10.1093/mnras/stab2426}}.

At the end of their life, massive stars explode as core-collapse supernova remnants (SNRs), hence the generated SNR blast waves will expand inside the complex wind-blown bubbles. Several SNRs such as Cas A (\myblue{\citet{2009A&A...503..495V, 2016ApJ...822...22O}}), SN 1987A (\myblue{\citet{1993ApJ...405..337B, 10.1093/mnras/sty1053}}), G$292+0.8$ (\myblue{\citet{Park_2002, Temim_2022}}) appear to evolve inside the wind-blown bubble. 
Evidently, the dynamics of the SNR blast waves will be regulated by the CSM parameters (\myblue{\citet{1989ApJ...344..332C, 1989A&A...215..347C, 2005ApJ...630..892D, 2007ApJ...667..226D,2021MNRAS.502.5340M,2022MNRAS.515..594M,2022MNRAS.515L..29M,2023MNRAS.519.5358V,2023MNRAS.521.5354M}}) and should not only differ from that for a uniform interstellar medium (ISM), but can also depend on the type of progenitor star. However, this is quite difficult to determine the type of progenitors for core-collapse SNRs from the morphology of SNRs as the observed asymmetries in SNRs can also be developed by the magnetic field and explosion mechanisms, for instance, \say{ear}-like structures observed in radio and X-ray emission have been interpreted as indicative of the interplay between the remnants and bipolar CSM, created by LBV progenitors \myblue{\citet{2021MNRAS.502..176C, 2021arXiv210801951U}}, whereas other authors attributed these structures to the orientation of the external magnetic field \myblue{\citep{Matt_2004, 10.1111/j.1365-2966.2005.08642.x, 2009MNRAS.395.1467P, 10.1111/j.1365-2966.2011.18239.x}} and explosion mechanisms, for example, jet-driven and single-lobed supernova \myblue{\cite{1999ApJ...524L.107K, hungerford2005gamma}}.\\
Supernova remnants (SNRs) are assumed to be major sources (\myblue{\citet{1934PNAS...20..254B,1994A&A...287..959D,2005ApJ...619..314B, 2013A&ARv..21...70B}}) of galactic cosmic rays (CRs), where CRs are considered to be accelerated by Diffusive Shock Acceleration (DSA) process (\myblue{\citet{1949PhRv...75.1169F, 1978MNRAS.182..147B, 1983RPPh...46..973D}}). In the core-collapse scenario, particle acceleration along with emission (\myblue{\citet{2013MNRAS.434.3368D}}) should be influenced by the interactions between SNR and wind-blown bubble. Particle acceleration during the expansion of the SNR through the wind bubble was studied assuming Bohm diffusion in \myblue{\citep{1988ApJ...333L..65V, 1989A&A...215..399B, 2000A&A...357..283B}}. Recently, \myblue{\citet{2022ApJ...926..140S, 2022ApJ...936...26K}} used simplified flow profiles to investigate the influence of an ambient medium shaped by Red Super Giant (RSG) and Wolf-Rayet (WR) winds on the non-thermal emission from the remnants. 

In our previous study (\myblue{\citet{2022A&A...661A.128D}}), the spectral evolution and emission morphology for SNR-CSM interaction were elaborately investigated using a realistic representation of CSM for the stellar track of a $60\,M_{\sun}$ star and for Bohm-like diffusion for energetic particles. Although in this study, the complete hydrodynamic evolution of CSM throughout the lifetime of the star has been considered, the treatment of magnetic turbulence was not self-consistent. The impact of the self-generated magnetic field amplification on the maximum attainable particle energies and the resulting softer spectra at higher energies during later times were intricately explored in \myblue{\citet{2016A&A...593A..20B, 2020A&A...634A..59B}} for type-Ia SNR. 

Probing the particle acceleration and non-thermal emission in SNR within wind bubbles by the combined influence of the CSM and the CR-streaming instabilities and their impact on CR diffusion would be desirable.
In this paper, we explore the spectral evolution of accelerated particles at forward shocks of SNRs with $20\,M_{\sun}$, and $60\,M_{\sun}$ progenitors, during their expansion inside the wind-blown bubbles, considering the time-dependent evolution of self-generated magnetic turbulence. The $20\,M_{\sun}$ star evolves through a RSG phase as the post-main sequence (MS) stage, whereas the $60\,_{\sun}$ star has LBV and WR phases after MS. Our study demonstrates the following aspects
\begin{itemize}

    \item The difference in spectral shapes arising from the morphological dissimilarity of the CSM of both SNRs.
    \item In both scenarios, the softening of particle spectra at higher energies during the later stages of evolution, on account of the weak driving of magnetic turbulence and the escape of high-energy particles from the remnants.
  
    \item The temporal evolution of the spectra for different emission channels as well as emission from the escaped particles around the remnants.
      \item The evolution of the morphology for the different energy bands during the lifetime of remnants and its dependence on the structure of the CSM.
\end{itemize}

\section{Numerical methods}

In this section, we introduce the numerical methods applied in the presented study. We have modelled the diffusive shock acceleration (DSA) at SNR forward shock in test-particle approximation by combining the hydrodynamic evolution of SNR inside CSM, large-scale magnetic field evolution in addition to the time-dependent treatment of the magnetic turbulence, and finally the solution for CR transport equation. We have numerically solved the particle acceleration and hydrodynamics, respectively, with \textit{RATPaC} (Radiation Acceleration Transport Parallel Code) \myblue{\citep{2012APh....35..300T,2013A&A...552A.102T, 2020A&A...634A..59B, 2018A&A...618A.155S}} and the \textit{PLUTO} code \myblue{\citep{2007ApJS..170..228M,vaidya_apj_865_2018}}. 

\subsection{Hydrodynamics}

\label{subsec:2.1}
The Euler hydrodynamic equations including an energy source/sink term, considering the magnetic field as dynamically unimportant, can be described as :
\begin{linenomath*}
\begin{equation}\label{eq:1}
   { \frac{\partial}{\partial t} \,   
\begin{pmatrix}
\rho \\
\bf m\\
E
\end{pmatrix}}
+ \nabla
 \,
\begin{pmatrix}
\rho\bf u \\
\textbf{m} \textbf {u} + P \bf I\\
(E+P)\bf u
\end{pmatrix}
^T =   
\,
\begin{pmatrix}
0 \\
0\\
S
\end{pmatrix}
\end{equation}
\end{linenomath*}
\begin{linenomath*}
\begin{equation}\label{eq:2}
    \frac {\rho \textbf {u}^2}{2}+\frac{P}{\gamma-1} = E; \quad \gamma = \frac{5}{3}
\end{equation}
\end{linenomath*}
where $\rho$, \textbf {u},\textbf{ m}, P, E, $S$ are the mass density, flow velocity, momentum density, thermal pressure, total
energy density, and source/sink term to include the optically-thin cooling and radiative heating for the construction of CSM at the pre-supernova stage, respectively. \textbf{I} is the unit tensor.  
\begin{figure}
\centering
\includegraphics[width=0.5\textwidth]{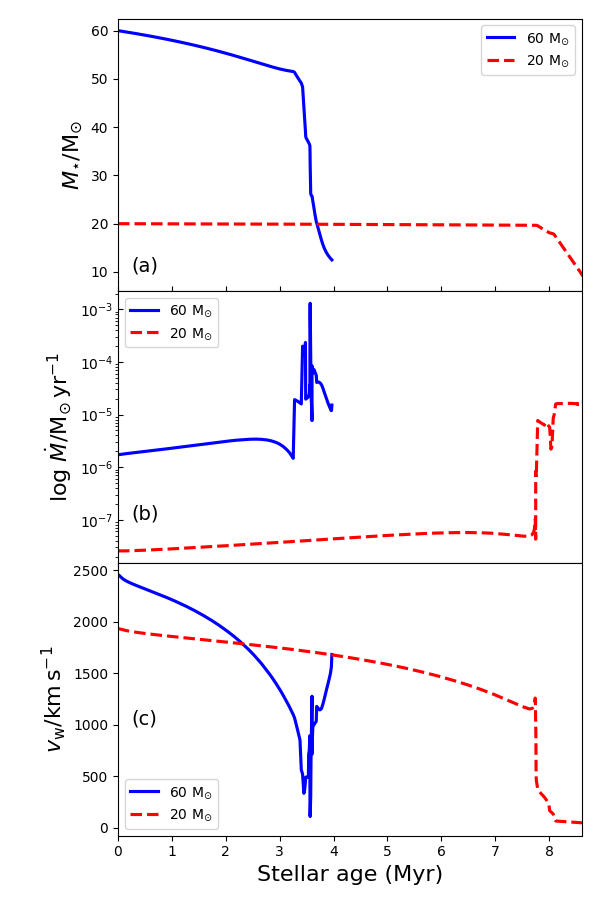}
\caption{{\textbf{Evolution of stellar mass in panel (a), mass-loss rate in panel (b), and wind velocity in panel (c) as a function of stellar age in $\mathrm{Myr}$ during different evolutionary stages of $20\,M_{\sun}$ and $60\,M_{\sun}$ stars shown in red dashed line and blue solid lines, respectively.} We used these parameters as the initial condition in hydrodynamic simulation for constructing the corresponding CSM at the pre-supernova stage at 8.64 million years for $20\,M_{\sun}$ star and 3.95 million years for $60\,M_{\sun}$ star. These stellar parameters for $20\,M_{\sun}$ and $60\,M_{\sun}$ stars are taken from \myblue{\citet{2012A&A...537A.146E, 2014A&A...564A..30G}}, respectively.}} 
\label{fig: Figure stellarevolution}
\end{figure}

\subsubsection{Construction of CSM at pre-supernova stage}\label{subsec:2.1.1}

We start with constructing the wind-blown bubble around the progenitor star with $20\,M_{\sun}$ at solar metallicity (Z = 0.014) from the zero-age main sequence (ZAMS) to the pre-supernova stage. \myblue{\citet{2022A&A...661A.128D}} already simulated the wind bubble for $60\,M_{\sun}$ star, and we use the same methodology also for the $20\,M_{\sun}$ progenitor. Hence, we will only provide a synopsis here.

We have performed hydrodynamic simulations with \textit{PLUTO} code in 1-D spherical symmetry with 50000 grid points of uniform spacing out to $R_{\mathrm{max}}=150\, \mathrm{pc}$. To initialise the simulation, the radially symmetric spherical supersonic stellar wind has been injected into a small spherical region of radius $0.06\,\mathrm{pc}$ at the origin, using the stellar evolutionary tracks for $20\,M_{\sun}$ star (\myblue{\citet{2012A&A...537A.146E}}), and $60M_{\odot}$ star (\myblue{\citet{2014A&A...564A..30G}}). The interstellar medium is assumed to have a constant number density, $n_\mathrm{ISM}=1\, \mathrm{cm^{-3}}$. The stellar wind density, $\rho_{\mathrm{w}}$, can be expressed as: 
\begin{linenomath*}
\begin{equation}\label{eq:3}
\rho_{\mathrm{wind}} = \frac{\dot{M}(t)}{4\pi r^{2} V_{\mathrm{w}}(t)}\ ,
\end{equation}
\end{linenomath*}

\begin{figure*}
\centering
\includegraphics[width=9.0cm, height=8.0cm]{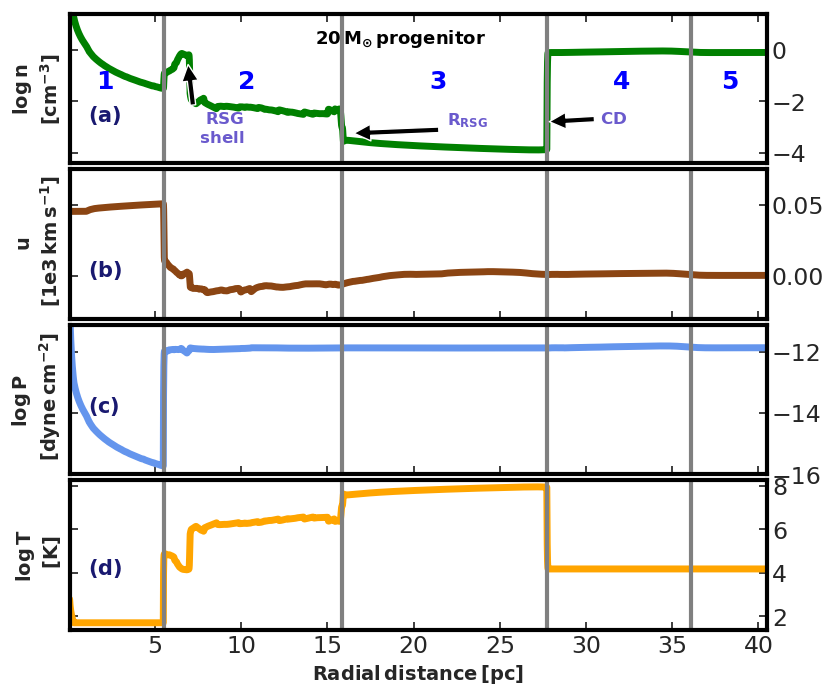}
\includegraphics[width=9.0cm, height=8.0cm]{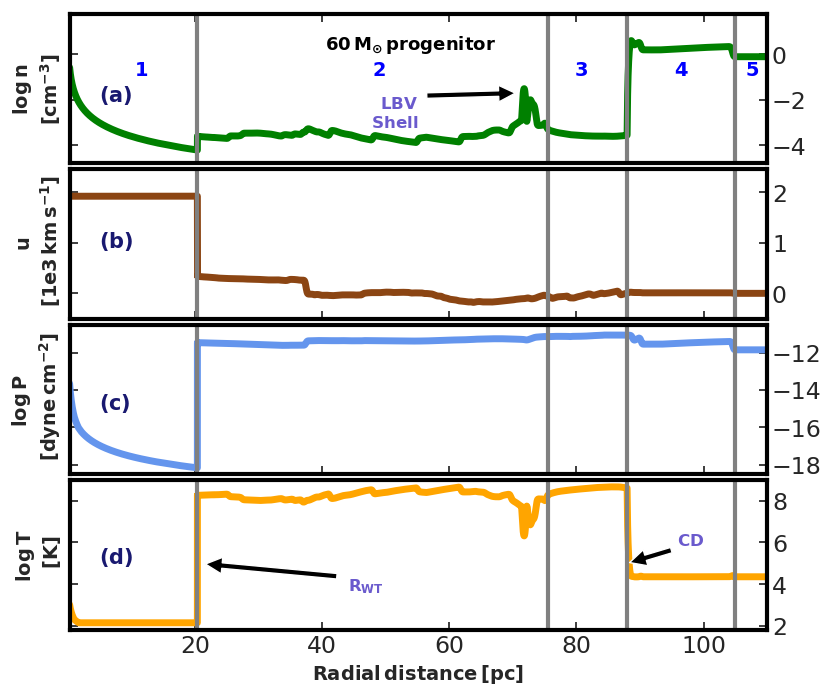}
\caption{\textbf{Pre-supernova CSM profiles of the number density, $n$, in panel (a), the flow speed, $u$, in panel (b), the thermal pressure, $P$, in panel (c), and the temperature, $T$, in panel (d), at 8.64 million years after ZAMS stage for the $\mathbf{20\,M_{\sun}}$ progenitor on the left and at 3.95 million years for the $\mathbf{60\,M_{\sun}}$ progenitor after ZAMS stage on the right. The right panels are reproduced from \myblue{\citet{2022A&A...661A.128D}.}}\\
Vertical grey lines mark the boundaries of specific regions. For the
$\mathbf{20\,M_{\sun}}$ progenitor these are the free RSG stellar wind (region 1), piled-up RSG wind (region 2), shocked MS wind (region 3), shocked ISM (region 4), and ISM (region 5). The RSG shell indicates the accumulation of decelerated, freely expanding RSG wind, $R_{\mathrm{RSG}}$ denotes the transition between RSG and MS wind, and CD represents the contact discontinuity between shocked wind and ISM.\\
For the $\mathbf{60\,M_{\sun}}$ progenitor we distinguish the free stellar wind (region 1), the shocked LBV and WR wind (region 2), shocked MS wind (region 3), shocked interstellar medium (ISM) (region 4), and ambient ISM (region 5). $\mathrm{R_{WT}}$ is the radius of wind termination shock (WT), LBV shell denotes the dense shell created by the interaction between LBV wind and WR wind, and CD represents the contact discontinuity between shocked wind and shocked ISM.
}
\label{fig: Figure 1}
\end{figure*}

where r is the radial coordinate, and $\dot{M}$ and $V_{\mathrm{w}}$ represent the time-dependent mass-loss rate and the wind velocity, respectively. The wind parameters for the $20\,M_{\sun}$ star have been derived from the Geneva library of stellar models \myblue{\citet{2012A&A...537A.146E}} and for the $60\,M_{\sun}$ star, wind properties have been taken from \myblue{\citet{2014A&A...564A..30G}}.

Figure \ref{fig: Figure stellarevolution} depicts the stellar mass in panel (a), mass-loss rate in panel (b), wind velocity in panel (c) during stellar evolution at different stages that are used as initial conditions in the hydrodynamic simulation for CSM. The CSM evolution has been simulated from ZAMS to the pre-supernova phase, over $8.64\, \mathrm {million\, years}$ for the $20\,M_{\sun}$ star and over $3.95\, \mathrm {million\, years}$ for the $60\,M_{\sun}$ star. The $20\,M_{\sun}$ star evolves through the MS phase for approximately 6.3 million years followed by the RSG stage and the $60\,M_{\sun}$ star evolves through the MS stage for approximately 3 million years followed by LBV phase with the span of approximately 0.2 million years and WR phase.
For modelling the evolution of the wind bubble, the second-order Runge-Kutta method has been applied to integrate equations \eqref{eq:1} and \eqref{eq:2}, using the Harten-Lax-Van Leer approximate Riemann Solver (hll) and finite volume methodology. Additionally, optically-thin cooling and radiative heating have been included through the source/sink term, $S = \Phi(T,\rho)$, following the cooling and heating laws,
\begin{linenomath*}
\begin{equation}\label{eq:Heatcool}
\Phi(T,\rho) = n^2(\Gamma(T)-\Lambda(T))
\end{equation}
\end{linenomath*}
where $n$ is the particle number density, $\Gamma$ is the radiative heating term that represents 
the effect of stellar photons.  
In our adopted cooling curve, the cooling term $\it \Lambda$ is composed of the contributions from 
hydrogen (H), helium (He) and metals ($Z$), whose contributions come from \myblue{\citep{wiersma_mnras_393_2009}} 
for a plasma at solar helium abundance \myblue{\citep{asplund_araa_47_2009}}, and it governs the 
gas cooling at high temperatures. 
Additionally, for temperatures $T< 6\times 10^{4}\, \mathrm{K}$, 
a cooling term for H recombination is added following the case B energy-loss 
coefficient of \myblue{\citet{hummer_mnras_268_1994}}. Collisionally excited forbidden lines (mainly from O and C elements) are included as described in Eq. A9 of \myblue{\cite{henney2009radiation}}, using a relative abundance $O/H=4.89\times 10^{-4}$ in number density \myblue{\citep{asplund_araa_47_2009}}.
The heating rate, ${ \Gamma}$, mimics the ionization of recombining 
hydrogenic ions by photons from the photosphere, in which the free electrons receive the energy carried by an ionizing photon once 
the reionization potential of an H atom 
has been subtracted. In a statistical equilibrium, the rate follows from the recombination coefficient $\alpha_{\mathrm{rr}}^{\mathrm{B}}$ whose values are from the table 4.4 of \myblue{\citep{osterbrock_1989}}. 
The heating term, $\Gamma$, represents the photoelectric
heating of dust grains by the Galactic far-UV background. For $T \le 1000\,
\mathrm{K}$, we used equation \,C5 of \myblue{\citet{wolfire_apj_587_2003}}, and the electron-density profile, $n_{\rm
e}$, follows Eq.\,C3 of \myblue{\citet{wolfire_apj_587_2003}}. For $T>1000\, \mathrm{K}$ we
take the value of $n_{\rm e}$ interpolated from the CIE curve
by \myblue{\citet{wiersma_mnras_393_2009}}. 
This cooling strategy has been used in a series of papers 
\myblue{\citep{2014MNRAS.444.2754M, 2020MNRAS.493.3548M, 2023MNRAS.521.5354M}}.

\par Fig. \ref{fig: Figure 1} illustrates the structure of the pre-supernova CSM at 8.64 million years and 3.95 million years after ZAMS stage for $20\,M_\odot$ and $60\,M_\odot$ progenitors, respectively. The wind bubble generated by the $20\,M_{\sun}$ progenitor consists of dense RSG material which stretches until around $r\approx 15\,\mathrm{pc}$, and the deceleration of slow RSG wind against the back-flowing MS material causes the formation of the RSG shell. This characteristic structure of CSM arising from RSG-MS interaction has been noted before for a $35\,M_{\sun}$ star \myblue{\citep[][Sec. 3]{2007ApJ...667..226D}}, and likewise the absence of a wind termination shock in the RSG wind. 
\par The temperature of the CSM around the $60\,M_{\sun}$ progenitor is very hot in the entire shocked wind material, extending out to $r\approx 60\,\mathrm{pc}$, whereas the CSM around the lower-mass star shows high temperature only in the region of shocked MS wind. The wind bubble structure around the $60\,M_{\sun}$ star has been illustrated before in \myblue{\citep{2022A&A...661A.128D}} and is reproduced in Fig. \ref{fig: Figure 1} to ease the comparison.\\

\par{\textbf{Comparative discussion about wind bubble structure}}\\
The structure of wind bubbles shaped during the lifetime of $20\,M_\odot$ and $60\,M_\odot$ progenitors reflects the history of the mass-loss rate and stellar-wind velocity as shown in Fig. \ref{fig: Figure stellarevolution}. In literature, a myriad of studies exist concerning the wind bubble structure and the included physics in the numerical modelling. For instance, the properties of the ISM and the prescription of radiative heating and cooling play a significant role in shaping the wind bubbles, along with the stellar-wind parameters as shown for two different stars in Fig. \ref{fig: Figure 1}. As the cooling time scale of gas depends on $\rho^{-1}$, for a high ISM density, the cooling timescale is small which leads to rapid cooling of the gas. For instance, \textcolor{myblue}{\cite{2011ApJ...737..100T}} considered an ISM density that is approximately 100 times higher than that in this paper and therefore very rapid cooling generates a thin and dense shell, whereas for the bubbles illustrated in Fig. \ref{fig: Figure 1} a thin dense shell of shocked ISM does not form. Including only radiative cooling, \myblue{\cite{2007ApJ...667..226D}} obtained a thin shell of shocked ambient material in the simulation of wind bubble around $35\,M_\odot$ star. The absence of radiative heating of the gas permits the gas to cool down to very low temperature and hence the shell becomes confined by the thermal pressure of the ambient medium. The prescription of radiative heating-cooling as in Equation \ref{eq:Heatcool}, with a balance temperature of $10^4\,\mathrm{K}$ in the ISM, resists the cooling of shocked ambient gas in the shell and leads a thick shell of shocked material without any sharp piled-up of shocked ISM material. The shape of the shell made by shocked ISM should also depend on the temperature of the ambient medium as the thermal pressure therein opposes the expansion of the shell. We mention in passing that dense shells at radiative shocks can also be suppressed by magnetic pressure \myblue{\citep{2016MNRAS.456.2343P}}.\\
The structure of this shell also relies on the wind parameters, as with a higher stellar mass the velocity of stellar wind increases, and so does its ram pressure, which supports the formation of prominent shells. The lifetime of stars decreases with stellar mass and hence, the propagation of the shock front of the shell for more massive stars becomes limited by the available time, whereas for less massive stars, the shock front evolves until the pressure of the shell comes into equilibrium with the ambient thermal pressure, and the shell structure depends on the balance temperature in the simulation.\\ 
Therefore, the structure of the shell is quite parameter-dependent, in particular concerning the ambient medium and the prescription for heating-cooling, and so care must be exercised in a comparison of different studies.\\ In this present study the structure of the CSM beyond the contact discontinuity of the wind bubble has negligible impact on the particle acceleration by the corresponding SNR, as particle acceleration becomes inefficient when the remnant reaches the CSM, and hence we stop our simulation within $1\,\mathrm{pc}$ beyond the contact discontinuity.
 \subsubsection{Supernova ejecta profile}\label{subsec:2.1.2}
The density of supernova ejecta, $\rho_\mathrm{ej}(r)$, is modelled as \textbf{\myblue{\citep{1999ApJS..120..299T}}}
\begin{linenomath*}
\begin{equation}\label{eq:4}
     \begin{split}
        \rho_{\mathrm{ej}}(r) = \begin{cases}\rho_\mathrm{c},  \qquad\qquad\qquad\quad r \leq r_c\\[.5em]
                \rho_\mathrm{c}\left(\frac{r}{r_\mathrm{c}}\right)^{-{n}} \quad\qquad r_\mathrm{c} < r \leq R_{\mathrm{ej}}\ ,
                \end{cases}
    \end{split}
\end{equation}
\end{linenomath*}
where $\mathrm{n} = 9$, as conventionally used for core-collapse explosions. The velocity profile for ejecta $u_\mathrm{ej}(r)$ follows homologous expansion,
\begin{linenomath*}
\begin{equation}\label{eq:5}
       u_\mathrm{ej} =\frac{r}{T_\mathrm{SN}}\ ,
\end{equation}
\end{linenomath*}
at the starting time of the hydrodynamic simulation, $T_\mathrm{SN}= 3\,\mathrm{years}$. The initial ejecta temperature is set to $10^4$K.
\par The expressions for $r_\mathrm{c}$ and $\rho_\mathrm{c}$ as a function of the ejecta mass, $M_\mathrm{ej}$, and explosion energy, $E_\mathrm{ej}$ will be,
\begin{linenomath*}
\begin{equation}\label{eq:6}
       r_\mathrm{c} = \left(\frac{10E_{\mathrm{ej}}}{3M_{\mathrm{ej}}}\,\frac{n-5}{n-3}\,\frac{n-3x^{3-n}}{n-5x^{5-n}}\right)^{1/2}T_{\mathrm{SN}}\ ,
\end{equation}
\end{linenomath*}
\begin{linenomath*}
\begin{equation}\label{eq:7}
      \rho_\mathrm{c} = \frac{M_{\mathrm{ej}}}{4\pi r_\mathrm{c}^3}{3(n-3)}\left(n-3x^{3-n}\right)^{-1}\ ,
\end{equation}
\end{linenomath*}
where $R_{\mathrm{ej}} = xr_\mathrm{c}$ and $x = 2.5$ \footnote{{There is a typo in the expression of $\rho_\mathrm{c}$ in Equation 7 of \myblue{\cite{2022A&A...661A.128D}}.}} Here, $x$ is a free parameter that determines the ratio between the flat and steep parts of the ejecta density profile and its choice is motivated by the comparability of the ejecta and ambient densities at $R_\mathrm{ej}$ and numerical stability of simulations. The choice of $x$ has a negligible effect on the dynamics of the shock which was thoroughly tested through the comparison of numeric simulations with analytic solutions \myblue{\citet{1982ApJ...258..790C, 1999ApJS..120..299T}}.
Further, in comparison to \myblue{\citet{1982ApJ...258..790C}} where the ejecta distribution extends to infinity, the truncation at $R_{\mathrm{ej}}$ provides a limit to the ejecta speed and feedback from pushing away the CSM that originally resided very close to the star. 
The explosion energy is fully transferred to ejecta, $E_\mathrm{ej}=10^{51}\ \mathrm{erg}$. The ejecta mass, $M_\mathrm{ej}$, for the $20\,M_{\sun}$ star is $3.25\,M_{\sun}$, and  $M_\mathrm{ej}= 11.75\,{M_{\sun}}$ for the $60\,M_{\sun}$ star, assuming a left-over compact object on the neutron-star mass scale.
\begin{linenomath*}
\begin{equation}\label{eq:8}
M_\mathrm{ej} = M_{\star} - \int_{t_{\mathrm{t_{ZAMS}}}}^{t_{\mathrm{preSN}}} \dot{M}(t)\mbox dt- M_{\mathrm{Compact Object}}(1.4M_{\sun})
\end{equation}
\end{linenomath*}
To launch the supernova explosions, the supernova ejecta profiles have been inserted in the pre-calculated pre-supernova CSM profiles, covering the inner 56 to 108 grid points, depending on the progenitor type and also on the ejecta mass. Then equations \eqref{eq:1} and \eqref{eq:2} have been solved in the absence of heating or cooling ($S=0$) using a Harten-Lax-Van Leer approximate Riemann Solver that employs the middle contact discontinuity (hllc), finite-volume methodology, and a second-order Runge-Kutta method. The numerical simulations of the supernova remnant with the \textit{PLUTO} code have been carried out in 1-D spherical symmetry with a spatial resolution of about $ 0.0004\,\mathrm{pc}$, about a factor 7.5 finer than that used for the wind bubble prior to the supernova explosion, using linear interpolation of the pre-supernova CSM grid.  
The proper solution of the cosmic-ray transport equation near the shock requires an exquisite spatial resolution, otherwise the cosmic-ray precursor is not resolved and the velocity jump at the shock is poorly reconstructed. The higher resolution of the grid describing the gas flow also helps in defining a sharp shock in the inhomogeneous spatial grid that we use for the cosmic-ray transport equation (cf. section~\ref{subsec:2.3}). To ensure a sharp transition in hydrodynamic parameters - density, velocity, pressure and temperature from upstream to downstream values at the shock, the hydro data from \textit{PLUTO} code needs resharpening. Resharpening is done by blinding hydro data for density, velocity, pressure and temperature at 4 bins upstream and downstream of the shock and replacing these values with the linear extrapolation using further downstream and upstream values \myblue{\cite[Chapter~4] {Brose:2020acu}}. These resharpened profiles are used for further necessary calculations in the simulation.
\subsection{Magnetic field}
\label{subsec:2.2}
The total magnetic field strength in the entire system is given by
\begin{linenomath*}
\begin{equation}\label{eq:9}
B_\mathrm{tot} = \sqrt{B_0^2+B_\mathrm{turb}^2}\ ,
\end{equation}
\end{linenomath*}
where $B_{0}$ and $B_\mathrm{turb}$ are the large-scale and turbulent magnetic field, respectively.
\subsubsection{Large-scale field profile}\label{subsec:2.2.1}
Simulating the CSM magnetic field for the entire lifetime of progenitors by solving magneto-hydrodynamic (MHD) equations is out of the scope of this paper. Therefore, we have parametrised the magnetic field in the CSM around $20\,M_{\sun}$ star, similarly as we constructed the CSM magnetic field for the $60\,M_{\sun}$ progenitor \myblue{\citep{2022A&A...661A.128D}}.
The magnetic field in the stellar wind of a rotating star will be toroidal except for very close to the stellar surface. In the equatorial plane of rotation, it can be expressed as \myblue{\citep{1998ApJ...505..910I, 1999IAUS..193..325G,1994ApJ...421..225C}},

\begin{linenomath*}
    \begin{equation}\label{eq:10}
       B_{\phi} 
       =  B_{\star} \frac{u_\mathrm{rot}R_{\star}}{u_\mathrm{wind}r} =\frac{B_{\star,0}R_{\star}}{r}\quad\qquad r >> R_{\star}\ ,
    \end{equation}
    \end{linenomath*}
where $B_{\star}$ and $R_{\star}$ are the stellar surface magnetic field and radius, respectively, $u_\mathrm{rot}$ and $u_\mathrm{wind}$ represent the surface rotational velocity in the equatorial plane and the radial wind speed, respectively.     
A $20\,M_{\sun}$ star evolves through MS and RSG phases. At the RSG phase, the star reaches a very large size up to a few hundred times $R_{\sun}$, but becomes a slower rotator (\myblue{\citet{RevModPhys.84.25}}), and the wind speed is $20-50\,\mathrm{km\,s^{-1}}$. Measurements by (\myblue{\citet{2010A&A...516L...2A, 2017A&A...603A.129T}}) suggest that a RSG has a weak surface field with $1-10\,\mathrm{G}$, and therefore we chose $B_{\star,0}(R_{\star}/R_{\sun})=750\,\mathrm{G}$. Equation \ref{eq:10} then defines the field strength for both free RSG wind and piled-up RSG wind, region 1 and 2 as marked in Fig. \ref{fig: Figure 1}. At $R_\mathrm{RSG}=16.3\,\mathrm{pc}\sim 7\times10^8\,R_{\sun}$, there is the transition to MS wind, which has a very uncertain magnetic field. We assume a decrease in magnetic field strength by a factor of $3$ following $B\propto\sqrt{\rho}$ and the density jump at $R_\mathrm{RSG}$. Throughout the MS wind the magnetic field is supposed to have a constant strength. In the ISM, the field strength has been chosen as $4\,\mu\mathrm{G}$ to provide a super-Alfvenic flow to the shell, marked as region 4 in Fig. \ref{fig: Figure 1}. The field in the shell has been calculated assuming flux conservation \myblue{\citep{2015A&A...584A..49V}}. In total, the magnetic field strength ($B_\mathrm{0,\,20\,M_{\sun}}$) in the different regions mentioned in Fig. \ref{fig: Figure 1} of the wind bubble created by $20\,M_{\sun}$ star can be expressed as,

\begin{linenomath*}
\begin{equation}\label{eq:11}
B_\mathrm{0,\,20\,M_{\sun}} = \begin{cases}(1.07\ \mathrm{\mu G})\frac{R_{\mathrm{RSG}}}{r} & \mathrm{regions}\, 1\,\& \,2\\[.5em]
       0.35\ \mathrm{\mu G} \qquad &  \mathrm{region}\,3\\[.5em]
       4.68\ \mathrm{\mu G}  \qquad &  \mathrm{region}\, 4\\[.5em]
       4.0\ \mathrm{\mu G} \qquad &   \mathrm{region}\, 5\ .
\end{cases}
\end{equation}
\end{linenomath*}
For the $60\,M_{\sun}$ star we used the same methodology as in \myblue{\citet{2022A&A...661A.128D}}, but slightly changed the field strength from those quoted there to maintain the magnetic field as dynamically unimportant. Hence, the magnetic field in the different regions of the CSM that is formed by a $60\,M_{\sun}$ progenitor, can be written as,
\begin{linenomath*}
\begin{equation}\label{eq:12}
B_\mathrm{0,\,60\,M_{\sun}} = \begin{cases}(0.63\ \mathrm{\mu G})\frac{R_{\mathrm{WT}}}{r} & \mathrm{region}\, 1\\[.5em]
       2.52\ \mathrm{\mu G} \qquad &  \mathrm{regions}\,2 \,\& \,3\\[.5em]
       14.8\ \mathrm{\mu G}  \qquad &  \mathrm{region}\, 4\\[.5em]
       4.3\ \mathrm{\mu G} \qquad &   \mathrm{region}\, 5\ .
\end{cases}
\end{equation}
\end{linenomath*}
We have determined the magnetic field initially in the supernova ejecta following \myblue{\citep[][Sec.3]{2013A&A...552A.102T}}.
Finally, the time-dependent computation of frozen-in large-scale magnetic field, transported passively with the hydrodynamic flow is given by the induction equation in 1D spherical symmetry \myblue{\citep{2013A&A...552A.102T}},
\begin{linenomath*}
\begin{equation}\label{eq:13}
    \frac{\partial \mathbf{B_0}}{\partial t} = \mathbf{\nabla}\times(\mathbf{u}\times \mathbf{B_0})\ .
\end{equation}
\end{linenomath*}

This method mimics MHD for negligible magnetic pressure.
\subsubsection{Magnetic turbulence}\label{subsec:2.2.3}
The temporal and spatial evolution of magnetic turbulence spectrum can be described by a continuity equation for magnetic spectral energy density per logarithmic bandwidth, $E_w(r, k, t)$ (\myblue{\citet{2016A&A...593A..20B}}), 
\begin{linenomath*}
 \begin{equation}\label{eq:15}
    \frac{\partial E_{w}}{\partial t} = -\nabla\cdot(\mathbf{u}E_{w})-k \frac{\partial}{\partial k}(k^2D_k\frac{\partial}{\partial k}\frac{E_{w}}{k^3})+2(\Gamma_g-\Gamma_d)E_w
\end{equation}
 \end{linenomath*}
where $k$ denotes the wavenumber, $D_{k}$ represents the diffusion coefficient in wavenumber space describing cascading, and $\Gamma_g$, $\Gamma_d$ are growth and damping rates, respectively. This transport equation for the magnetic turbulence spectrum has been solved in 1D spherical symmetry, considering Alfvén waves as scattering centres for CRs.\\
\myblue{\citet{2009MNRAS.392.1591A}} suggested that although non-resonant CR streaming instabilities are likely the dominant way of magnetic field amplification during the free expansion phases and early Sedov-Taylor phase of SNR, when the SNR shock is relatively fast, at later times resonant modes provide efficient amplification. It is to be noted that non-resonant amplification is a very complex issue, and some of the commonly quoted concepts may be misperceptions. The magnetic field in the free wind, that the shock passes through early in the evolution, is mostly perpendicular to the shock normal, whereas the usual calculations of the growth rate for resonant (\myblue{\citep{1975MNRAS.172..557S, 1978MNRAS.182..147B}}) and non-resonant \myblue{\citep{2004MNRAS.353..550B}} modes is performed for streaming parallel to the magnetic field. Here we may have oblique streaming, but that implies that even fewer growth times are available before the plasma passes through the shock. Already for parallel shocks, one expects only very few growth times, i.e. few exponential growth cycles \myblue{\citep{2008ApJ...684.1174N}}. This is one of the reasons why the peak amplitude of the non-resonant mode is generally less than naively estimated \myblue{\citep{2021ApJ...921..121P}}. The entire process, and hence its spatial profile, is highly nonlinear. It would be important to understand the cascading properties of the mode and the dependence on the wave spectrum of the CR diffusion coefficient, but that is beyond the scope of this paper. In the literature, cascading is often neglected and Bohm diffusion is simply assumed \textcolor{myblue}{\citep[e.g.][]{2008ApJ...678..939Z, 2021A&A...650A..62C}}. Here, we enhance by a factor $A=10$ the growth term for the resonant streaming instability,
\begin{linenomath*}
 \begin{equation}\label{eq:16}
  \Gamma_g = A \frac{v_Ap^2v}{3E_w}\abs*{\frac{\partial N}{\partial r}}
\end{equation}
\end{linenomath*}
where $v$ and $p$ are the particle velocity and momentum respectively, $v_A$ is the Alfvén speed, $N$ is the differential number density of CRs. The value $A=10$ agrees with the observations of historical SNRs (\myblue{\citet{2021A&A...654A.139B}}) and estimates of the growth rates operating at the early stages of CR-acceleration \myblue{\citep{2018MNRAS.479.4470M}}. 

The spectral energy transfer process from small wavenumber scale to large wavenumber scale through turbulence cascading can be empirically described as a diffusion process in wavenumber space with coefficient (\myblue{\citet{https://doi.org/10.1029/JA095iA09p14881, 2002cra..book.....S}})
\begin{linenomath*}
 \begin{equation}\label{eq:17}
 D_{k} 
 = k^3\,v_A\,\sqrt{\frac{E_{w}}{2B_0^2}}\ .
\end{equation}
\end{linenomath*}
This form of the diffusion coefficient conforms with Kolmogorov scaling in the case of pure cascading from large scales, $E_w \propto k^{-2/3}$. Furthermore, since $v_A \propto B_\mathrm{tot}$ and $B_\mathrm{turb}^2=4\pi \smallint d\ln k\ E_w(k)$, $D_{k}$ depends more sensitively on $E_w$, when the turbulent field is amplified beyond the amplitude of the large-scale field, 
\begin{linenomath*}
\begin{equation}\label{eq:18}
D_k = \begin{cases} \sqrt{E_w}\qquad & \mathrm{for} \qquad E_w \ll B_0^2/8\pi\\[.5em]
       E_w \qquad & \mathrm{for} \qquad E_w \gg B_0^2/8\pi .
\end{cases}
\end{equation}
\end{linenomath*}

\subsubsection{Diffusion coefficient}\label{subsec:2.2.4}
The diffusion coefficient governs the rate of particle acceleration, the maximum attainable energy of the particles \myblue{\citep{1983A&A...125..249L, 2010MNRAS.406.2633S}}, and the spatial distribution of the accelerated particles in the remnant. Since we have considered Alfvén waves as the scattering centres for CRs, the resonance condition is,
\begin{linenomath*}
 \begin{equation}\label{eq:14}
   k_\mathrm{res} = \frac{\vert q\vert\,B_0}{pc}
\end{equation}
 \end{linenomath*}
where $k_\mathrm{res}$ represents the resonant wavenumber, and $q$ is the particle charge. Then, the diffusion coefficient for CRs coupled to $E_w$ can be described as (\myblue{\citet{1978MNRAS.182..147B, 1987PhR...154....1B}}),

\begin{linenomath*}
 \begin{equation}\label{eq:20}
 D_\mathrm{r} = \frac{4v}{3\pi}r_g\frac{U_B}{E_w}
\end{equation}
\end{linenomath*}
where $U_{B}$ is the energy density of the large-scale magnetic field and $r_g$ represents the gyro-radius of particles in the total magnetic field ($B_\mathrm{tot}$).
The growth of Alfvén waves, damping, spectral energy transfer through cascading, and the spatial transport of waves suggest that the magnetic turbulence spectra should exhibit a complicated shape rather than a featureless and flat spectrum described by the Bohm-like diffusion coefficient. In addition, the time-dependent calculation of $E_w$, following equation \ref{eq:20}, provides the self-consistent diffusion coefficient, hence resulting in a more realistic picture of CR acceleration in SNR, in comparison to the simplified Bohm-like diffusion elucidated in our previous study with core-collapse SNR \myblue{\citep{2022A&A...661A.128D}}. 

\begin{figure*}
\centering
\includegraphics[width=9cm, height=8cm]{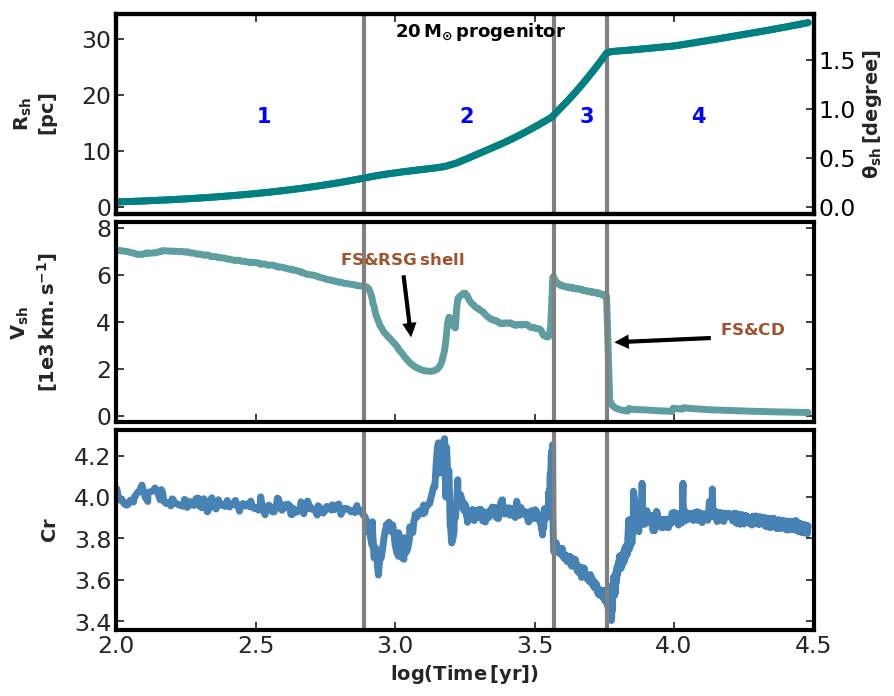}
\includegraphics[width=9cm, height=8cm]{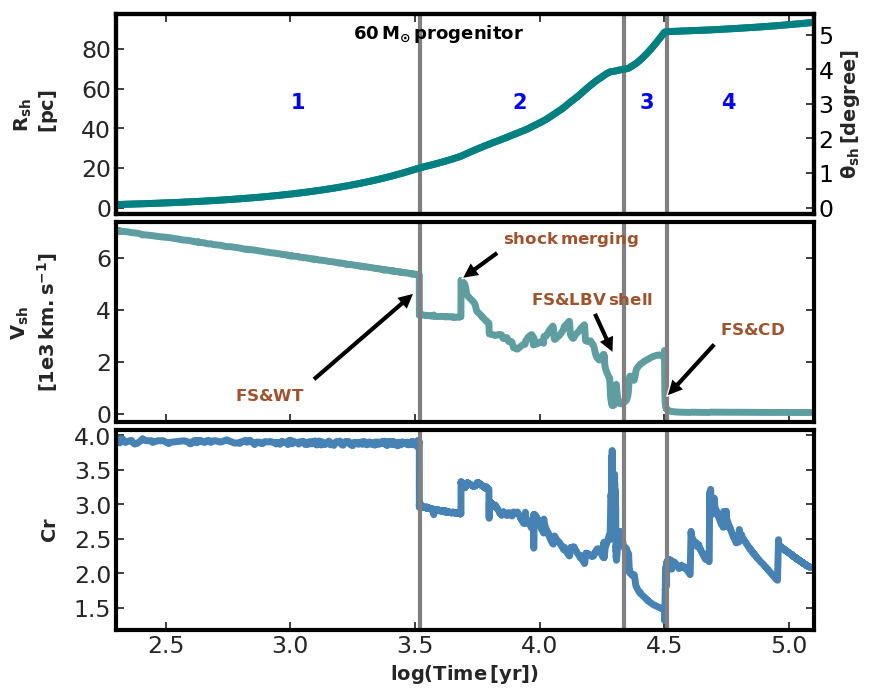}
\caption{\textbf{Temporal evolution of the forward shock parameters:} the
radius, $R_\mathrm{sh}$, the speed, $V_\mathrm{sh}$, and the sub-shock compression ratio, $\mathrm{Cr}$. In the upper panel, we provide the angular
scale, $\theta_\mathrm{sh}$, for an SNR located $1000 \mathrm{pc}$ away. We also mark the interactions of the forward shocks with discontinuities in the wind bubbles of the two types of progenitor stars.}
\label{fig: Figure 2}
\end{figure*}
As an initial condition, we assume a magnetic turbulence spectrum that yields a diffusion coefficient a factor of 100 lower than that for Galactic propagation of cosmic rays \myblue{\citep{Trotta_2011}}, 
\begin{linenomath*}
 \begin{equation}\label{eq:21}
  D_0 = \left(10^{27}\ \mathrm{cm}^2\,\mathrm{s}^{-1}\right) \left(\frac{pc}{10\mathrm{GeV}}\right)^{\frac{1}{3}}\left(\frac{B}{3\mu G}\right)^{-{\frac{1}{3}}}\ .
\end{equation}
\end{linenomath*}

\subsection{Particle acceleration}
\label{subsec:2.3}
The time-dependent transport equation for the differential number density of CRs, $N(p)$, can be written as, 
\begin{linenomath*}
 \begin{equation}\label{eq:22}
    \frac{\partial N}{\partial t} = \nabla (D_\mathrm{r}\nabla N - \mathbf{u}N) - \frac{\partial }{\partial p}\left(\dot pN - \frac{\nabla\cdot\mathbf{u}}{3}Np\right) +Q
 \end{equation}
 \end{linenomath*}
where $\dot p$ corresponds to  the energy loss rate for synchrotron and inverse Compton losses of electrons \myblue{\citep{2021A&A...654A.139B}}, and $Q$ denotes the source term \myblue{\citep{1975MNRAS.172..557S}}. 

This equation has been solved in test-particle approximation for spherical symmetry in \textit{RATPaC}, applying implicit finite-difference algorithms implemented in the FiPy package \myblue{\citep{2009CSE....11c...6G}}. In this paper, the non-linear shock modification by CR pressure is negligible and ignored, because the cosmic-ray pressure has always constrained below $10\%$ of the shock ram pressure \myblue{\citep{2010ApJ...721..886K}}. We have solved 
equation \ref{eq:22} on an inhomogeneous spatial grid, $r^\prime=r/R_\mathrm{sh}(t)$ and this co-ordinate transformation provides excellent spatial resolution near the shock  \myblue{\citep{2020A&A...634A..59B, 2022A&A...661A.128D}},\textbf{
\begin{linenomath*}
\begin{equation}
r^\prime-1 = (x_\ast -1)^3 \ .
\end{equation}
\end{linenomath*}
}
where $R_\mathrm{sh}$ is the shock radius.

\subsection{Injection of particles} 
\label{subsubsec:2.3.1}
The source term, $Q$, in the transport equation, can be expressed as
\begin{linenomath*}
 \begin{equation}\label{eq:23}
  Q = \eta_\mathrm{inj} n_\mathrm{u} (V_{\mathrm{sh}}-u_\mathrm{u})\delta(R-R_{\mathrm{sh}})\delta(p-p_{\mathrm{inj}})\ ,
\end{equation}
\end{linenomath*}
where $\eta_\mathrm{inj}$ is the injection efficiency, $n_\mathrm{u}$ and $u_\mathrm{u}$ are the upstream plasma number density and velocity in observer's frame, respectively, $V_{\mathrm{sh}}$ is the shock velocity in the observer's frame, and $p_{\mathrm{inj}} = \xi p_{\mathrm{th}}=\xi\sqrt{2m k_BT_d}$ , represents the momentum of injected particles. The injection efficiency is defined following the thermal leakage model \myblue{\citep{2005MNRAS.361..907B}}, 

\begin{linenomath*}
 \begin{equation}\label{eq:24Rsub}
  \eta_\mathrm{inj} = \frac{4}{3\pi^{1/2}}(R_\mathrm{sub}-1)\xi^3\exp{(-\xi^2)}
 \end{equation}
 \end{linenomath*}
where $R_\mathrm{sub}$ represents the sub-shock compression ratio, the ratio of upstream and downstream flow speed in the shock rest frame where $V_\mathrm{sh}$ is calculated by mass flux through the shock in the observer's frame and $u_\mathrm{u}$ and downstream plasma velocity in observer's frame ($u_\mathrm{d}$) are calculated at the immediate upstream and downstream at the shock using resharpened flow velocity profile. We have used $\xi=4.2$ in our simulations, which is consistent with the observed radiation flux from SN1006 \myblue{\citep[][Appendix A]{2021A&A...654A.139B}} and conforms with the test-particle limit. Furthermore, as we have injected electrons and protons at the same $\xi$, the electron-to-proton ratio at higher energy can be determined by their mass ratio, $K_{ep}\sim \sqrt{m_e/m_p}$ (\textcolor{myblue}{\cite{pohl1993predictive})}.

\section{Results}
We follow the evolution of the SNRs of the $20\,M_{\sun}$, and $60\,M_{\sun}$ progenitor
stars for 30000 years and 110000 years, respectively. We discuss the behaviour of parameters for the SNR forward shock along with the spectra for accelerated particles. Furthermore, we compare the spectra, specifically at later times, obtained from the self-consistent calculations with explicit turbulence transport with those derived for Bohm-like diffusion, illustrated in \myblue{\citep{2022A&A...661A.128D}}. Additionally, we present the spectra and morphology of non-thermal emissions from the SNRs.
\subsection{Shock parameters}
\label{subsec:3.1}
Fig. \ref{fig: Figure 2} illustrates the forward shock parameters for both SNRs.
\subsubsection{SNR with \texorpdfstring{${20\,M_{\sun}}$ progenitor}{TEXT}}  
During the expansion through free RSG stellar wind region, the speed of the forward shock gradually decreases from $\mathrm{6300\,km\,s^{-1}}$ to $\mathrm{5300\,km\,s^{-1}}$, and the sub-shock compression ratio is approximately 4. When the forward shock starts interacting with the dense RSG shell, the shock speed falls to about $\mathrm{2000\,km\,s^{-1}}$, and the sub-shock compression ratio slightly diverges from the value 4 and reaches 3.7. For brief moments around 1500 years and 3600 years the compression ratio becomes approximately 4.2, when the forward shock propagates along a steeply declining density, once after reaching the peak of RSG shell and then at the transition from the piled-up RSG wind to the shocked MS wind. During this period, the upstream density decreases more rapidly in comparison to the downstream density which consequently results in a slightly higher compression ratio at the forward shock. Thus, the fluctuations in compression ratio reflect interactions with discontinuities in the CSM. The hot MS wind material in region 3 causes reduction in the sonic Mach number, $M_\mathrm{s}$, and consequently, the compression ratio becomes approximately 3.5.

\subsubsection{SNR with \texorpdfstring{${60\,M_{\sun}}$ progenitor}{TEXT}}
The evolution of the forward shock parameters has been described in \myblue{\citet[][Sec. 3.1]{2022A&A...661A.128D}}. The most prominent feature is that the sub-shock compression ratio reaches approximately 1.5 while the SNR expands through the hot shocked wind.\\
The SNR shock interactions with any structures in CSM give rise to reflected shocks that propagate through the downstream of SNR forward shock and eventually hit other structures present there and reflected back in the outward direction. These shocks collide with SNR forward shock and produce instantaneous spikes in the shock velocity, as shown in Fig. \ref{fig: Figure 2}. This feature has already been described elaborately in \myblue{\cite{2022A&A...661A.128D}}.
\subsection{Particle acceleration and escape}
\label{subsec:3.2}
\begin{figure}
\centering
\includegraphics[width=\columnwidth]{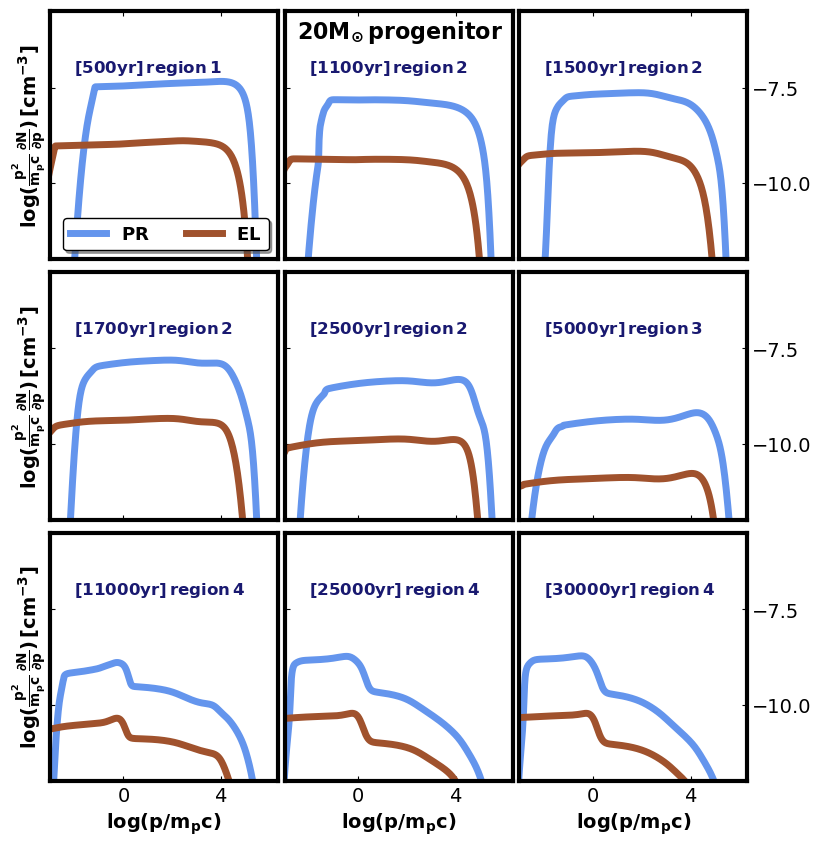}
\caption{\textbf{Proton (PR) and electron (EL) spectra volume-averaged downstream of
the forward shock at different regions of the corresponding wind bubble for $\mathbf{20\,M_{\sun}}$ progenitor (cf. Fig.~\ref{fig: Figure 1}).}}
\label{fig: Figure 3}
\end{figure}

\begin{figure}
\centering
\includegraphics[width=\columnwidth]{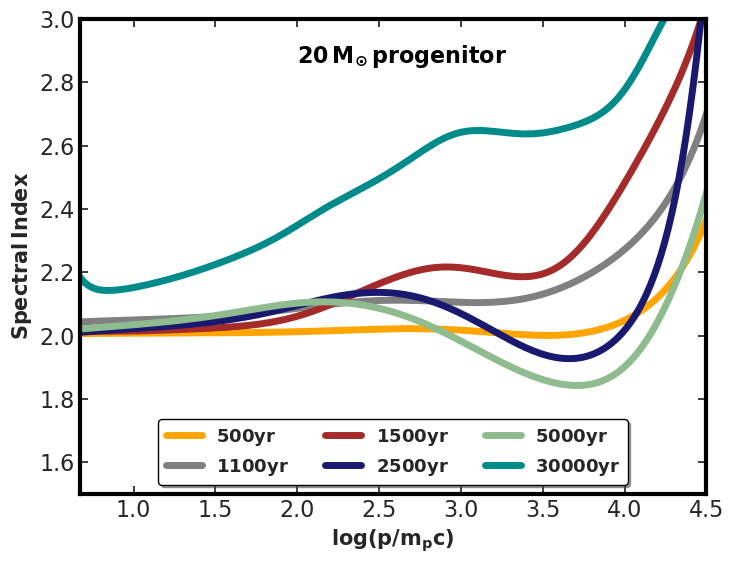}
\caption{\textbf{Variation of the spectral index for downstream protons at different ages with momentum.}}
\label{fig: Figure 4}
\end{figure}

To understand the spectral evolution of accelerated particles, the volume-averaged downstream spectra for protons and electrons are illustrated in Fig. \ref{fig: Figure 3} and Fig. \ref{fig: Figure 6} for the two types of SNRs, and the corresponding spectral indices are shown in Fig. \ref{fig: Figure 4} and Fig. \ref{fig: Figure 7}, respectively. The selected ages correspond to the forward shock passing through the different regions of wind bubbles. The earlier study of type-Ia SNR with self-consistent amplification of Alfvénic turbulence \myblue{\citep{2020A&A...634A..59B}} indicated that the spectral shape is controlled by the transport of turbulence and Alfvénic diffusion. The continuous deceleration of the shock decreases the CR flux and hence reduces the CR gradient, leading to a weaker driving of turbulence. For the core-collapse SNRs discussed here, the resulting particle spectra should get affected both by the complicated hydrodynamics of the wind bubbles and by the dynamics of the self-consistent turbulence. Further, the profiles resulting from the passive transport of CSM magnetic field (cf. sec. \ref{subsec:2.2.1}) can also influence the particle acceleration.   \\
\begin{figure}
\centering
\includegraphics[width=\columnwidth]{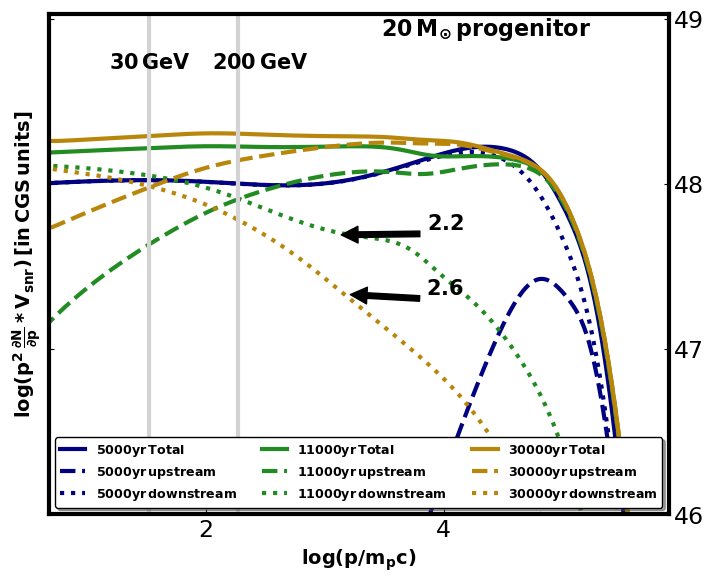}
\caption{\textbf{Proton number-spectra at later times}. Solid lines represent the total spectra and dashed, and dotted lines indicate the forward shock upstream, and the downstream spectra, respectively. The gray vertical lines at $\mathrm{200\,GeV}$ and $\mathrm{30\,GeV}$ denote the escape energies at 11000 years and 30000 years, respectively. The arrows point out the energy bands with spectral indices of 2.2 and 2.6.}
\label{fig: Figure 5}
\end{figure}
\subsubsection{SNR with \texorpdfstring{${20\,M_{\sun}}$ progenitor}{TEXT}}
{The particle acceleration in the different regions of the wind bubble is as follows-}\\
\textbf{Free RSG wind}-During the propagation of SNR forward shock inside this region, the compression ratio becomes 4 and so, the particle spectral index is around $2$, here shown at 500 years in Fig. \ref{fig: Figure 3} and Fig. \ref{fig: Figure 4}. At the end of the evolution inside this region, the maximum attainable energy for protons reaches $20\,\mathrm{TeV}$. \\
\textbf{Piled-up RSG wind}- During the interaction between the forward shock and dense RSG shell, the maximum attainable energy of protons slightly decreases to $6\,\mathrm{TeV}$, on account of the decrease in shock speed. During this time, illustrated at 1100 years and 1500 years, spectral softening is observed, and the spectral index reaches $ 2.2$ at energies above $100\,\mathrm{GeV}$ for protons, but the spectral index returns to $2$ when the forward shock has emerged from the RSG shell. This brief spectral softening reflects the variation of the sub-shock compression ratio depicted in Fig. \ref{fig: Figure 2}. After climbing the RSG shell, the shock speed again increases, and so does the maximum energy of particles. During this time the compression ratio becomes 4.2, which results in spectral hardening to an index of $1.9$ at higher energies, shown at 2500 years. Further, the collision between the piled-up RSG wind and the forward shock gives rise to an inward-moving reflected shock which merges with the SNR reverse shock and propagates towards the interior with a speed of $\sim 3000\,\mathrm{km.s^{-1}}$. We do not inject particles at this shock, but this inward-moving shock can re-accelerate energetic particles. As we display the total volume-averaged particle spectra downstream of the forward shock, at higher energies the hardest contribution should dominate \myblue{\citep{1972ApJ...174..253B, 2001A&A...377.1056B}}.\\
\textbf{Shocked MS wind}- In this region, the forward shock passes through hot MS wind material, and consequently the forward shock has a reduced compression ratio and produces relatively soft particle spectra, but these particles are so few that volume-averaged over the downstream region we do not observe a spectral softening (e.g. at 5000 years in Fig. \ref{fig: Figure 3}). The high temperature in the shocked MS wind also provides a high injection momentum, and the volume-averaged spectrum is dominated by previously accelerated particles at all energies. The freshly accelerated particles in the MS wind cannot penetrate deep into the interior, on account of the strong magnetic field in the piled-up RSG wind.\\
\textbf{Shocked ISM}-After passing through the contact discontinuity between shocked MS wind and shock ISM, the forward shock becomes inefficient at accelerating particles up to very high energy. In the shocked ISM region, the maximum attainable energy for protons falls from $20\,\mathrm{TeV}$ to $50\,\mathrm{GeV}$ at the end of our simulation. The forward shock encounters dense and cool material in this region which lowers the injection momentum and increases the injection rate into DSA. But the forward shock is too weak to accelerate these particles to very high energy. Hence, a prominent spectral break near $1\,\mathrm{GeV}$ is noticeable in Fig. \ref{fig: Figure 3} at 11000 years and later. Later, the multiple merging of fast shocks with the weak forward shock increases the acceleration efficiency, and the step-like spectral feature emerges at slightly higher energies. The electron spectra roughly follow those of protons, except for an additional softening arising from the radiative cooling.  \\
In our time-dependent treatment of the transport of CRs and Alfvénic turbulence, the driving of turbulence diminishes at later stages of SNR evolution on account of the decrease in CR pressure gradient (\myblue{\citet{2020A&A...634A..59B}}). Therefore, the diffusion coefficient as well as the acceleration timescale (\myblue{\citet{1991MNRAS.251..340D}}) increase, hence particles with higher energy in the shock precursor are slow to return and participate in further shock acceleration. Hence, a break in the downstream particle spectra should appear at the currently achievable maximum energy, above which particles escape to the far-upstream region. After 30000 years this softening is clearly visible in Fig. \ref{fig: Figure 4}, where the spectral index reaches approximately 2.6 at high energy, starting from about 2.2 above $10\,\mathrm{GeV}$. 

Total production spectra of protons, including particles outside of the SNR, are depicted in Fig. \ref{fig: Figure 5} at different times. As we solve the CR transport equation out to $65\, R_\mathrm{sh}$, all particles reside inside the simulation domain. Therefore, integrating the particle spectra over the whole simulation domain yields the total CR production of the SNR, and that is spectrally harder than the component inside the remnant. At 5000 years, protons are accelerated up to $20\,\mathrm{TeV}$ energy while at 11000 years protons above $200\,\mathrm{GeV}$ are preferentially found in the upstream region, rendering the downstream spectra softer with the spectral index of $\sim 2.2$, shown in Fig. \ref{fig: Figure 5}. At a later stage, the transition energy shifts to $30\,\mathrm{GeV}$ and the spectral index reaches $\sim 2.6$ above this escape energy. At any time, the break appears at the energy up to which the particles can be currently accelerated. It reflects the interplay between the reduction in maximum energy and the escape of particles from the accelerator (\myblue{\citet{2010A&A...513A..17O, 2019MNRAS.490.4317C, 2020A&A...634A..59B}}).\\
\begin{figure}[!b]
\centering
\includegraphics[width=\columnwidth]{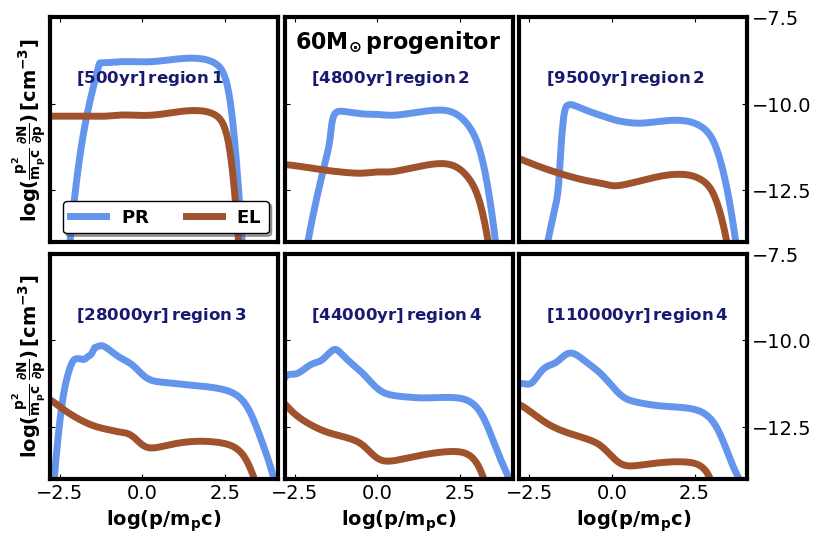}
\caption{\textbf{Proton (PR) and electron (EL) spectra volume-averaged downstream of
the forward shock at different times and regions of the corresponding wind bubble of the 60-$M_\odot$ progenitor (cf. Fig.~\ref{fig: Figure 1}).}}
\label{fig: Figure 6}
\end{figure}
\begin{figure}
\centering
\includegraphics[width=\columnwidth]{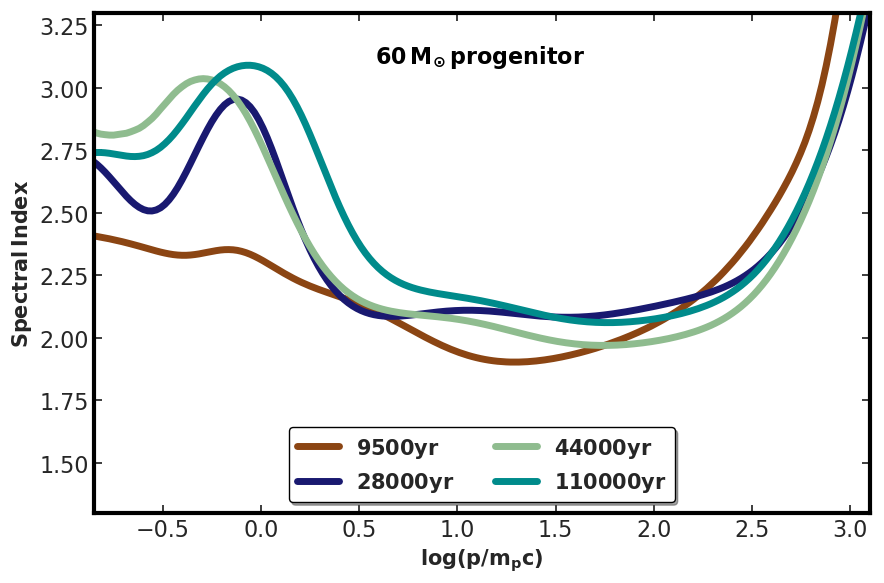}
\caption{\textbf{Variation of the spectral index for protons at
different ages with momentum.}}
\label{fig: Figure 7}
\end{figure}
\begin{figure}
\centering
\includegraphics[width=\columnwidth]{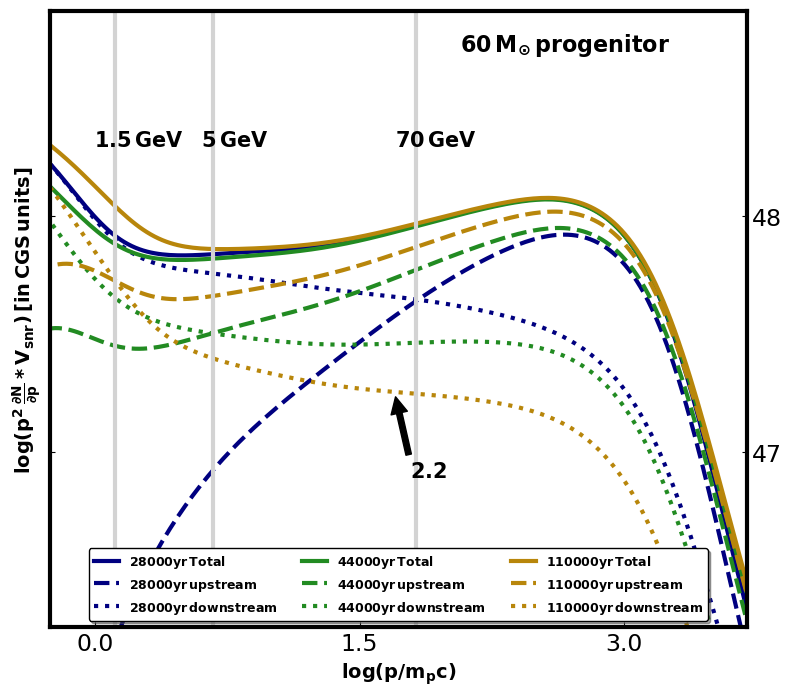}
\caption{\textbf{Proton number-spectra at later times}. Solid lines
represent the total spectra. The dashed and dotted lines indicate the spectrum upstream and downstream of the forward shock, respectively. The gray vertical lines at $\mathrm{70\,GeV}$, $\mathrm{5\,GeV}$, and $\mathrm{1.5\,GeV}$ denote the escape energies at 28000 years, 44000 years, and 110000 years, respectively. The arrow points out the energy bands with spectral index of 2.2.}
\label{fig: Figure 8}
\end{figure}

\textbf{SNR with $\mathbf{60\,M_{\sun}}$ progenitor}- \\
Particle acceleration by the SNR forward shock at different regions of the wind bubble was explored in our previous work \myblue{\citep{2022A&A...661A.128D}} for Bohm-scaling of the diffusion coefficient. Here we investigate the spectral modifications deriving from the explicit treatment of turbulence transport, considering the CR streaming instability and the growth, damping, and cascading of the waves. The propagation of the SNR forward shock through the hot shocked regions, region 2 and region 3 in Fig. \ref{fig: Figure 1}, leads to softened spectra below roughly $10\,\mathrm{GeV}$, but not as severe as for Bohm scaling, in particular at higher energies. Later, after 28000 years, the spectral index reaches 2.75 at lower energies, agreeing with the result for Bohm-like diffusion. Additionally, the introduction of turbulence adds non-linearity to the calculation of the diffusion coefficient. 
While the SNR shock expands through the shocked ISM, with the self-consistent turbulence model the spectral softening at higher energies arises from the evanescence of Alfvén waves. After 110000 years, the spectral index reaches approximately 2.2 above $\sim10\mathrm{GeV}$, illustrated in Fig. \ref{fig: Figure 8}, on account of the escape of particles from the remnant. Further, the proton production spectra shown in Fig. \ref{fig: Figure 8} suggest that the particles above $70\,\mathrm{GeV}$ 
\begin{figure}
\centering
\includegraphics[width=\columnwidth]{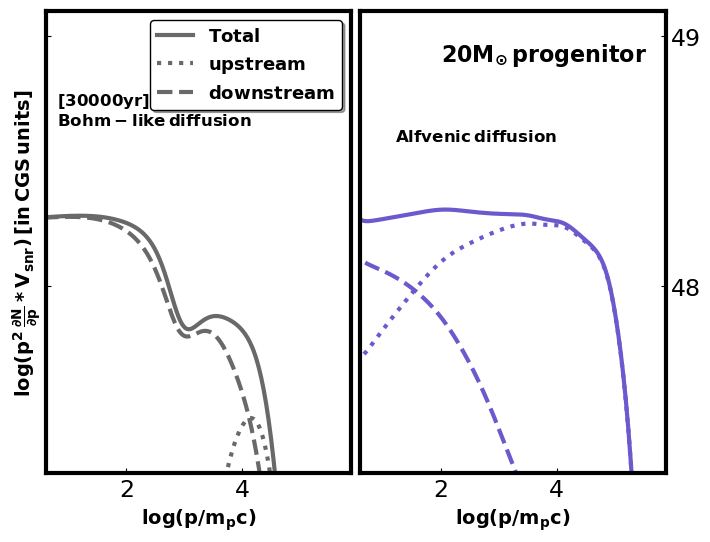}
\caption{\textbf{Proton number-spectra for Bohm-like (left panel) and self-consistent (right panel) diffusion.}}
\label{fig: Figure 9}
\end{figure}
start leaving the remnant already after 28000 years when the forward shock resides inside the shocked MS wind. Then, this typical energy of escaping CRs is reduced to about $1.5\,\mathrm{GeV}$ after 110000 years as a consequence of the inefficient confinement of high energy particles in the Alfvénic diffusion scenario.\\

At SNR with $20\,M_{\sun}$ progenitor the total proton production spectra have an index of $\sim 2.1-2.2$ above approximately $100\,\mathrm{GeV}$ energies at 11000 years and 30000 years. For the $60\,M_{\sun}$ progenitor, the index of the total proton production spectra is $\sim 1.9$ near the cut-off at later times. Both spectral shapes are quite different from the production index for type-Ia SNRs, about $2.4$ above $10\,\mathrm{GeV}$ \textcolor{myblue}{\citep{2020A&A...634A..59B}}. There the downstream spectra are shaped by the temporal decline of the maximum energy $E_\mathrm{max}$, to which the shock can accelerate \citep[see also][]{2019MNRAS.490.4317C}. In contrast, the downstream spectra at core-collapse SNRs also reflect the effect of the CSM hydrodynamics, besides the time evolution of turbulence spectra. For example, the SNR with $60\,M_{\sun}$ provides downstream spectra with index $\sim 1.9$ at higher energies, before the escape starts. Although at later times the downstream spectra become softer as the consequence of particle escape, the proton production spectra show a slight hardness near cut-off, shown in Fig. \ref{fig: Figure 8}.

Our results suggest that the cascading and decay of turbulence are crucial in the formation of soft particle spectra with spectral breaks for the older remnant. In contrast, with simple Bohm-like diffusion particles near the cut-off energy may escape the shock environment only at later stages, yielding insignificant spectral modifications, as shown in Fig. \ref{fig: Figure 9} for SNR with $20\,M_{\sun}$ progenitor as well as for $60\,M_{\sun}$ progenitor described by \textcolor{myblue}{\citet{2022A&A...661A.128D}}. We emphasise that the spectral shape with Bohm-like diffusion only reflects the influence of the CSM flow profiles, while in the Alfvénic scenario the spectra are a function of the hydrodynamic parameters and the properties of the turbulence. 
Our investigation on particle acceleration in SNRs with $20\,M_{\sun}$ 
and $60\,M_{\sun}$ progenitors illustrate the differences in spectral features arising from the environment of the SNRs. This is to be noted that the flow structure of wind bubbles becomes eventually very complex, on account of multiple reflected and transmitted shocks. We have solved the CR transport equation on a non-uniform grid centered on the forward shock to obtain the required fine resolution near the shock, described in sec. \ref{subsec:2.3}. Simultaneously refining the grid also for the other shocks located downstream is desirable but out of the scope of this paper. Further, acceleration of particles from very low energies is possible only for the forward shock, and so we inject the particles into DSA only there. Re-acceleration of pre-energised particles at the other shocks, in particular the reverse shock, is included though, and we have seen negligible effects of that in the particle spectra. The higher fraction of reaccelerated particles in \citet{2022A&A...661A.128D} is entirely caused by their choice of a Bohm-like diffusion coefficient.
We obtain softer spectra at high energies for the SNR with $20\,M_{\sun}$ progenitor, during the propagation of the SNR forward shock in the shocked ISM, while for the very massive progenitor the signature of softness is observed already when the SNR shock is in the shocked wind, on account of particle  escape. Here, the difference between a Bohm-like diffusion coefficient \citep{2022A&A...661A.128D} and the self-consistent treatment in this work becomes particularly evident, as the softness at low energies is present in both cases, due to the hot shocked wind, but at higher energies considerably softer spectra only develop due to particle escape.
The enhanced driving of turbulence in the scenario with the $20\,M_{\sun}$ progenitor arises from the higher normalisation of the CR spectrum (\myblue{\citet{2020A&A...634A..59B}}), that is caused by roughly four orders of magnitude higher density in the RSG wind. Additionally, the shock radius for the $20\,M_{\sun}$ progenitor is smaller than for the $60\,M_{\sun}$ progenitor, which enhances the gradient in the cosmic-ray distribution and hence the driving rate of turbulence.
Fig. \ref{fig: Figure 6} suggests that the maximum cut-off energy for protons is $E_{\mathrm{max}}\approx 0.5\,\mathrm{TeV}$ for the $60\,M_{\sun}$ progenitor, which is significantly less than that for the $20\,M_{\sun}$ progenitor, $E_{\mathrm{max}}\approx 50\,\mathrm{TeV}$. A contributing factor could be our estimation of ejecta mass, $M_\mathrm{ej}=11.75\,M_{\sun}$ for SNR with $60\,M_{\sun}$ progenitor, assuming the compact object as a neutron star. But, in reality, a black hole might be formed in the explosion of this very massive star, which would result in a lower ejecta mass. Then the ejecta speed and the shock speed would be higher, leading to a maximum energy of particles that is higher by a factor of a few, but not more, because the acceleration rate scales with $v_\mathrm{sh}^2$ and hence inversely with the ejecta mass for a given explosion energy.
In addition, the result from SNR with $20\,M_{\sun}$ progenitor is consistent with the estimation of maximum energy for type II progenitor discussed in \myblue{\citet{2020APh...12302492C}}. Further, the spectral index at high energies obtained for the SNR with $20\,M_{\sun}$ progenitor at later times is roughly comparable to those observationally deduced for IC443 and W44 (\myblue{\citep{2013Sci...339..807A}}). 
\subsection{Non-thermal emission}
\label{subsec:3.3}
The processes of non-thermal radiation include synchrotron emission, inverse Compton scattering, and the decay of neutral pions. In our results, the synchrotron cut-off energy reflects the evolution of the electron acceleration efficiency that depends on the magnetic field profile and hydrodynamics. 
In this paper we only consider the cosmic microwave background to derive the inverse Compton emission from the remnants. \myblue{\citet{2006ApJ...648L..29P}} estimated that the contribution of infrared and optical photon fields to inverse Compton emission is not
significant to that of CMB photons except for SNRs residing near the Galactic centre. Hence, we are aware that these additional photon fields might be needed to consider during the modelling of specific SNRs by taking into account the respective background photon fields.
The emission flux from different processes is calculated for remnants at $1 \mathrm{kpc}$ distance.
Fig. \ref{fig: Figure 11} and \ref{fig: Figure 16} illustrate the synchrotron spectra from the SNRs with $20\,M_\sun$ and $60\,M_\sun$ progenitors respectively, and Fig. \ref{fig: Figure 12}, \ref{fig: Figure 17} demonstrate the gamma-ray spectra from both the remnants. These illustrations depict the time evolution of the total emission spectra, as well as the part originating downstream of the SNR shock. Fig. \ref{fig: Figure 13} demonstrates as a function of time the radio flux at $5\,\mathrm{GHz}$, the non-thermal X-ray flux in the $0.1\,\mathrm{keV} - 10\,\mathrm{keV}$ range, as well as the gamma-ray emission at high energy (HE) ($0.1\,\mathrm{GeV}-100\,\mathrm{GeV}$) and very high energy (VHE) ($>1\,\mathrm{TeV}$) for the SNR with $20\,M_{\sun}$ progenitor.{We also calculate intensity \textbf{profiles} at characteristic times for synchrotron  (Fig. \ref{fig: Figure 14} and Fig. \ref{fig: Figure 18}) and gamma-ray emission (Fig. \ref{fig: Figure 15} and Fig. \ref{fig: Figure 19}), following the method described in \textcolor{myblue}{\citet{2013A&A...552A.102T}}. 
Although total X-ray emission may be largely thermal during the interaction of the remnant with dense molecular clouds \textcolor{myblue}{\cite{2021A&A...649A..14U}}, efficient CR acceleration may suppress the post-shock temperature and hence the thermal X-ray flux \textcolor{myblue}{\cite{2009A&A...496....1D}}. Calculating the thermal X-ray emission from SNRs is out of the scope of this study and only the non-thermal emission is presented.
\begin{figure}[b]
\centering
\includegraphics[width=\columnwidth]{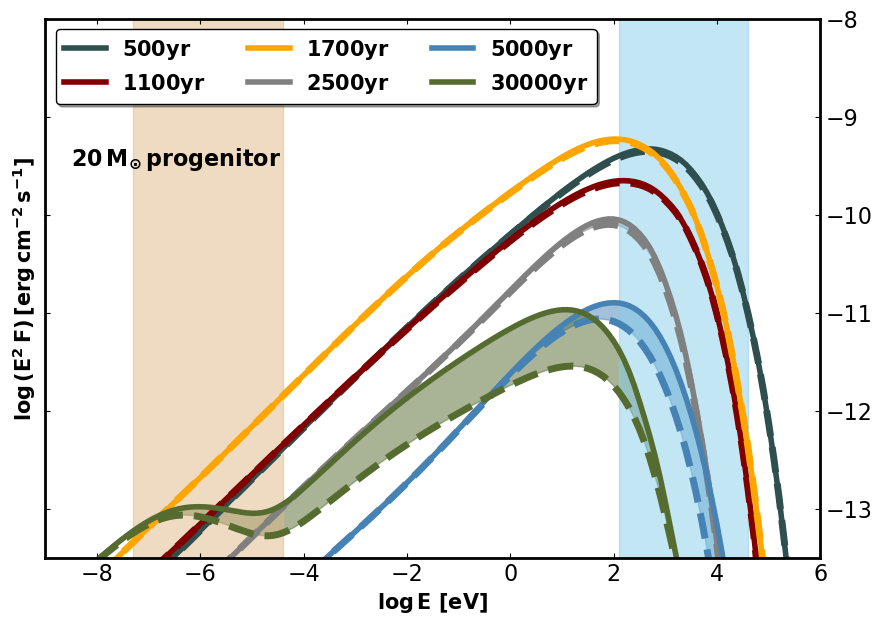}
\caption{\textbf{Synchrotron emission from the SNR at different ages.} The upper
boundaries of the shaded regions indicate the total emission from the remnant
whether the lower ones denote emission from the downstream of SNR forward shock. The brown band indicates the $50\,\mathrm{MHz}-10\,\mathrm{GHz}$ range and the blue band denotes $0.1\,\mathrm{keV}-40\,\mathrm{keV}$.}
\label{fig: Figure 11}
\end{figure}\\
\textbf{SNR with $20\,M_\sun$ progenitor}- Below the age of about $5000\,\mathrm{years}$, essentially the entire non-thermal emission is produced in the interior of the remnant.\\
\textbf{Free RSG wind}- At the early stages of evolution, the interaction of the SNR with the strong magnetic field in the free wind region, $B_{0}\propto 1/r$, combined with strong amplification, yields the considerable X-ray flux at this stage, shown at 500 years in Fig. \ref{fig: Figure 11}. The simulated non-thermal X-ray light curve shown in Fig. \ref{fig: Figure 13} indicates declining X-ray emission during the SNR evolution inside free RSG wind. Initially, the remnant expands through dense material, $\rho\propto r^{-2} $ in the free RSG wind, and consequently pion-decay emission dominates over inverse-Compton scattering for the first 200 years.
\begin{figure}[t]
\centering
\includegraphics[width=\columnwidth]{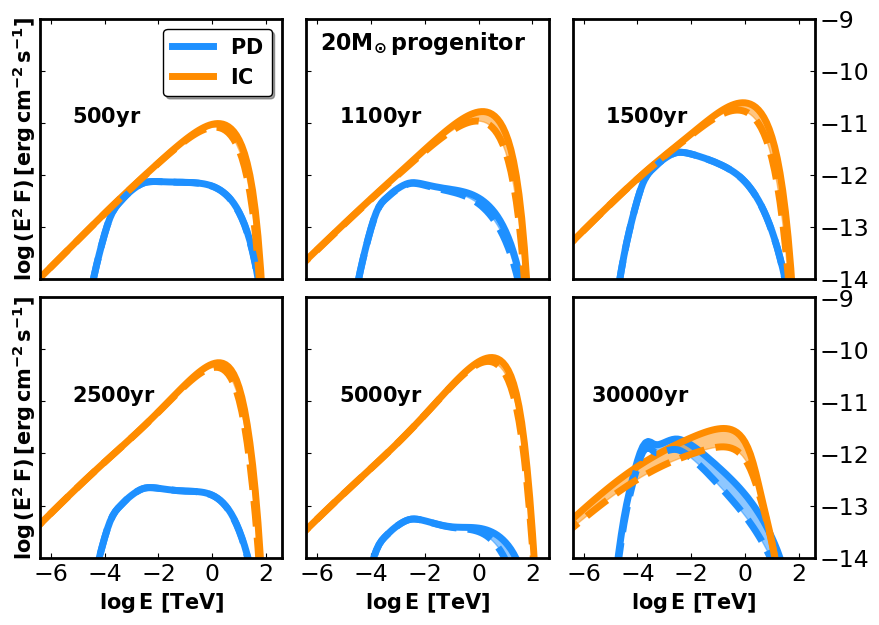}
\caption{\textbf{Gamma-ray emission by pion-decay (PD) and inverse Compton (IC) scattering at different ages.} The boundaries of the shaded regions indicate the total emission and that from the interior, as in Fig. \ref{fig: Figure 11}}
\label{fig: Figure 12}
\end{figure}
\begin{figure}[t]
\centering
\includegraphics[width=\columnwidth]{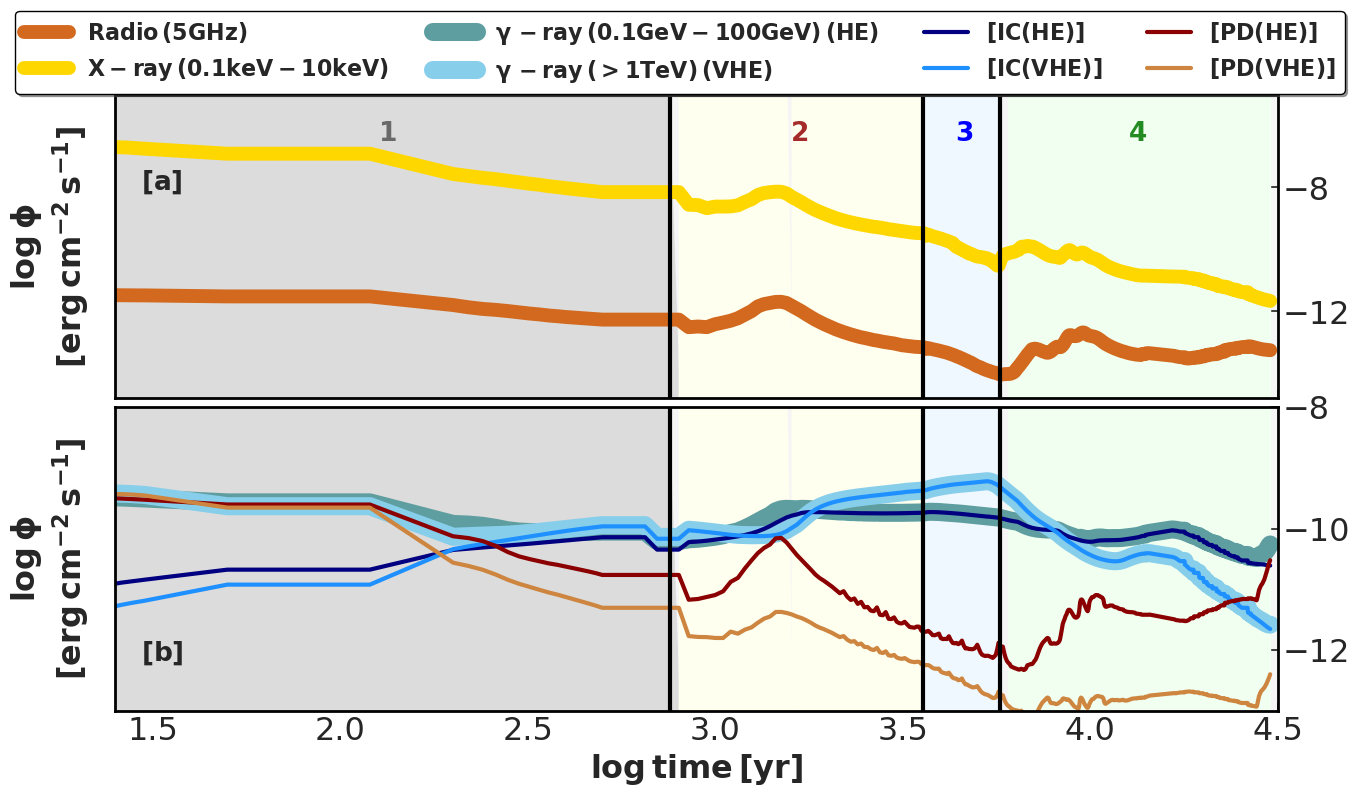}
\caption{\textbf{Evolution of energy flux ($\Phi$) during the lifetime of SNR for synchrotron emission and gamma-ray emission
at specific energy ranges} inside the different regions shown in Fig. \ref{fig: Figure 1} of wind-bubble formed by $20\,M_{\sun}$ progenitor. We specify that the shown X-ray flux only refers to the non-thermal X-ray.}
\label{fig: Figure 13}
\end{figure}
\begin{figure*}[h!]
\centering
\includegraphics[width=15.0cm, height=10cm]{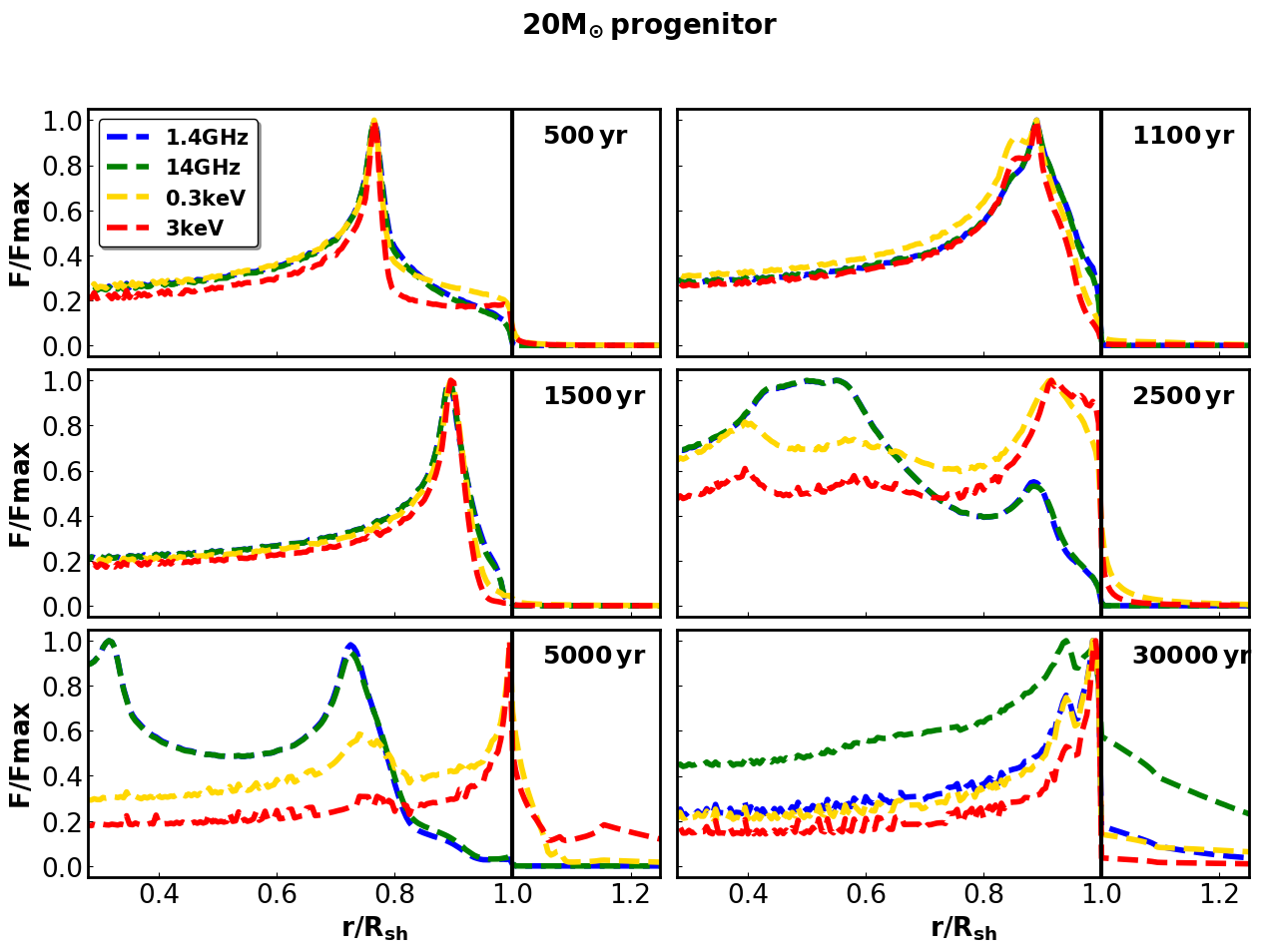}
\caption{\textbf{Normalised intensity profiles for synchrotron
emission.} For each specified profile, the intensity is normalised
to its peak value.}

\label{fig: Figure 14}
\end{figure*}
\begin{figure*}[h!]
\centering
\includegraphics[width=15.0cm, height=10cm]{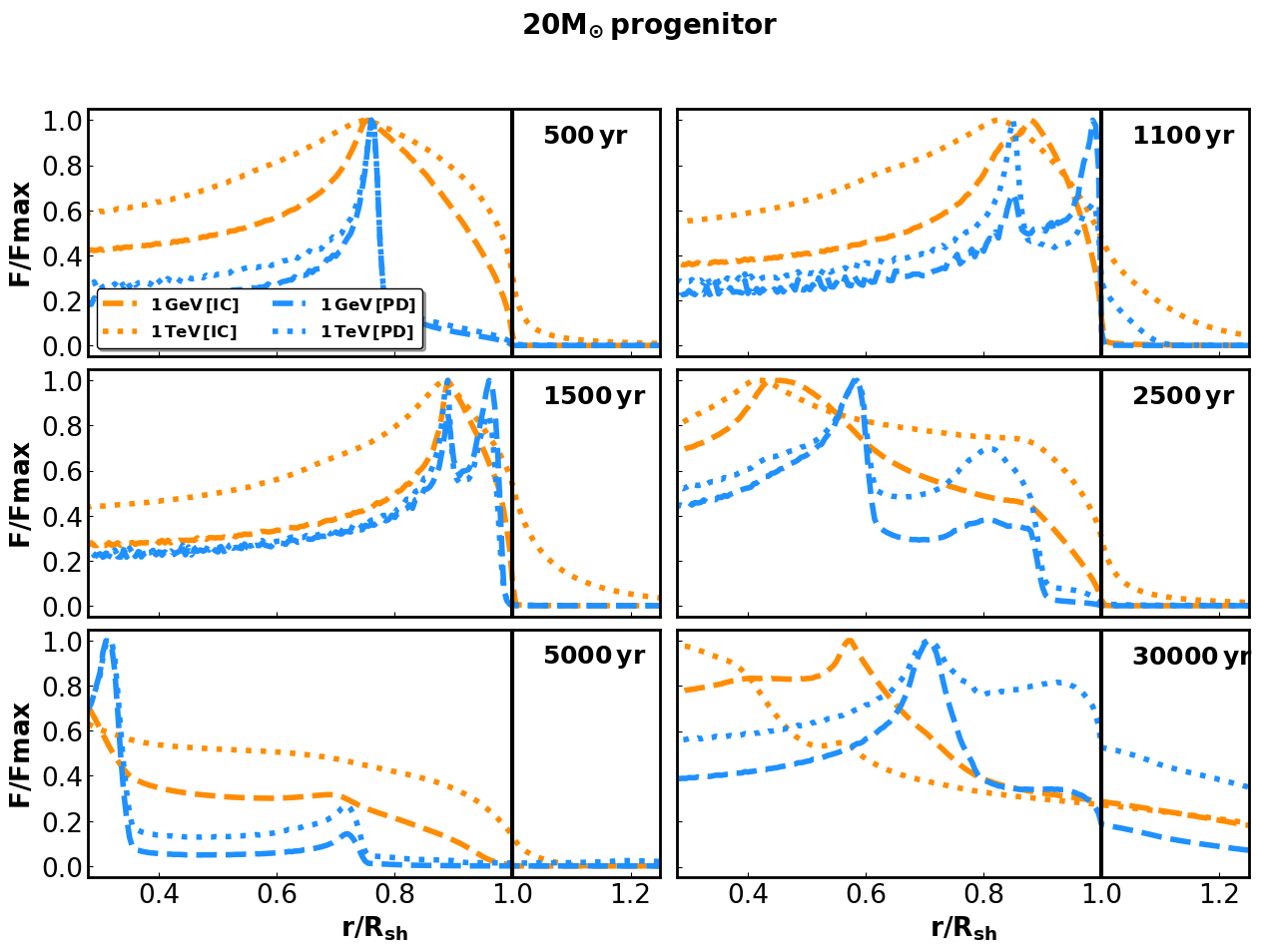}
\caption{\textbf{Normalised intensity profiles for gamma-ray
emission.} Emissions for pion-decay (PD) and inverse Compton emissions (IC)are shown. For each specified profile, the intensity is normalised to its peak value.}

\label{fig: Figure 15}
\end{figure*}
During this stage, the strongest magnetic field is found near the contact discontinuity between SNR forward and reverse shock \myblue{\citep[cf.][]{2004ApJ...609..785L}}. The turbulent magnetic field generated in the upstream and at the shock is eventually damped in the downstream region as a consequence of the weak driving of turbulence \myblue{\citep{2005ApJ...626L.101P}}. Hence, the contact discontinuity between forward and reverse shock of the SNR produces the highest intensity in the radio and X-ray band, shown in Fig. \ref{fig: Figure 14}. Further, Fig. \ref{fig: Figure 15} indicates that at this stage, the peak intensity of pion-decay and inverse Compton emission also coincides with the contact discontinuity between the forward and reverse shock. The interior of the remnant also appears inverse-Compton bright at $1\,\mathrm{TeV}$, because the high-energy electrons can penetrate the deep interior.\\
\textbf{Piled-up RSG wind}- During the collision of SNR with the RSG shell, at an age between $750\,\mathrm{years}$ and $1600\,\mathrm{years}$, the radio spectra start to soften to a spectral index $\alpha \approx 0.54$ ($S_{\nu}\propto \nu^{-\alpha}$) well in agreement with the low-energy spectral index of particles as seen in Fig. \ref{fig: Figure 4} but after the crossing of the RSG shell the radio spectral index resumes to be $\alpha = 0.5$.\\
In addition, the pion-decay flux enhances on account of the interaction with dense RSG shell. Further, during this interaction, the spectral index of pion-decay emission reaches $2.2$. It is important to note here, that the softness of the pion-decay emission is extending all the way to low gamma-ray energies, unlike the all-downstream particle spectrum, as the bulk of the hadronic emission originates from the shocked RSG-shell immediately downstream of the shock. As a result, the low-energy gamma-ray spectra appear even softer than the radio emission that is produced by electrons of comparable energy. 
This stage of SNR evolution may be comparable to Cas A which may be expanding inside the dense RSG wind \myblue{\citep{2003ApJ...593L..23C}}. Although the ambient CSM of Cas A may differ from that of a $20\,M_{\sun}$ progenitor, and the reverse shock of Cas A is an efficient accelerator (\myblue{\cite{2008ApJ...677L.105U}}), the spectral index for accelerated protons obtained in this study is comparable with that estimated by \myblue{\citet{2014A&A...563A..88S}}. The synchrotron flux in the radio and X-ray band as well as the hadronic emission flux reaches their maximum at this time when the forward shock crosses the dense RSG shell. For the next 200 years after that, the X-ray flux and HE pion-decay flux decrease at rates of $\sim 0.3\%/\mathrm{yr}$ and $\sim 0.7\%/\mathrm{yr}$, respectively, as a consequence of the declining density and the deceleration of the remnant. For Cas A, a reduction rate of $1.5\%/\mathrm{yr}$ was observed for non-thermal X-rays in the band $4.2\,\mathrm{keV}-6\,\mathrm{keV}$ (\textcolor{myblue}{\citet{Patnaude_2011}}). So, it is possible that the forward shock of Cas A is propagating through piled-up RSG wind, and the difference in density, age, and shock radius may reflect a much lower progenitor mass for Cas A than the $20\,M_{\sun}$ assumed here.

Further, the synchrotron morphology is centre-filled in this region. After the encounter with the RSG shell, the velocity of forward shock starts to increase, and the reverse shock and the contact discontinuity move towards the interior. After $2500$~years, two radio shells are visible, the inner shell at the contact discontinuity and the outer one at the shocked RSG shell.
Additionally, the brightest X-ray band is created near the forward shock. In the gamma-ray band, the brightest gamma-ray emission emanates from the region near the contact discontinuity between forward and reverse shock, and reverse shock. In reality, the intensity peak is likely smeared out, on account of the Rayleigh-Taylor instability of the contact discontinuity. Furthermore, at 2500 years the entire remnant looks bright in inverse Compton emission, specifically at $1\,\mathrm{TeV}$, as the energetic particles reach the reverse shock and may be re-energised there. \\

\textbf{Shocked MS wind}- At 5000 years, when the SNR forward shock is inside this region and about to collide with the contact discontinuity between shocked MS wind and shocked ISM, Fig. \ref{fig: Figure 11} indicates that the total X-ray synchrotron flux is slightly higher than that coming from the downstream region only. The intensity of X-ray emission at this time, shown in Fig. \ref{fig: Figure 14}, indicates that the upstream emission is generated near or at the wind bubble contact discontinuity, on account of the very strong magnetic field there, almost 15 times stronger than that in the shocked MS wind. At the same time, the highest radio intensity comes from the piled-up RSG wind behind the forward shock and also from the region inside of the contact discontinuity between the SNR forward and reverse shock. In this stage the pion-decay flux decreases on account of the lower density of the medium, and the VHE inverse-Compton emission flux dominates over the HE flux, as the maximum achievable energy for electrons increases on account of the increase in shock velocity by $2000\,\mathrm{km.s^{-1}}$ in this region.
The normalised intensity \textbf{profile} at 5000 years suggests that both the leptonic and hadronic gamma-ray emission emerges from deep downstream.\\
\textbf{Shocked ISM}- During the forward-shock passage through the shocked ISM, particles of higher energy can escape the remnant, as discussed in sec. \ref{subsec:3.2}, and a significant fraction of the synchrotron flux is produced in the upstream region, depicted at 30000 years in Fig. \ref{fig: Figure 11}. Additionally, the pion-decay emission flux enhances on account of the shock propagating in dense material. At this age, a fraction of the gamma-ray emission is produced around the remnant, on account of particle escape. The spectral index for pion-decay emission reflects the soft proton spectra, and it reaches $2.4-2.6$ above $10\,\mathrm{GeV}$. This kind of soft gamma-ray spectra is observed from SNRs for example IC443, W44, $G\,39.2-0.3$, etc., that are expanding in or near the dense molecular clouds (\myblue{\citep{10.1093/mnras/stt1318, 2014A&A...565A..74C, 10.1093/mnras/staa2045}}). The old remnant appears shell-like in pion-decay emission, whereas the inverse Compton emission is centre-filled, contrary to the synchrotron intensity \textbf{profile}. This morphology of old core-collapse remnant is consistent with that of old type-Ia SNRs \myblue{\citep{2021A&A...654A.139B}}.

\begin{figure}[b]
\centering
\includegraphics[width=\columnwidth]{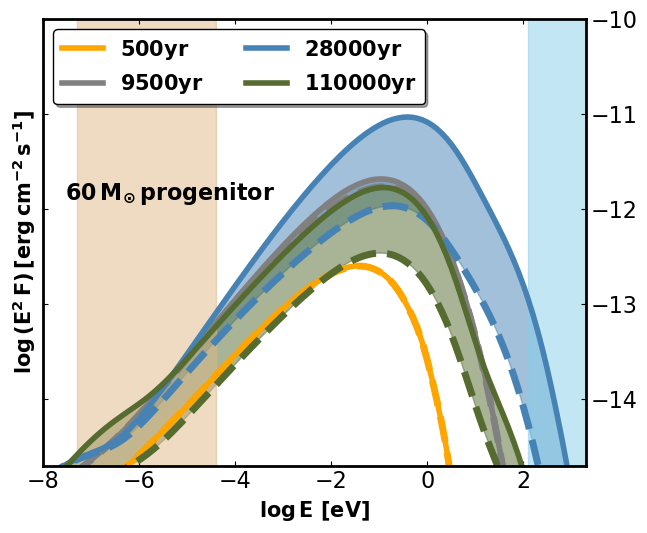}
\caption{\textbf{Synchrotron spectra from the SNR at different ages.}  The boundaries of the shaded regions indicate the total emission and that from the interior of the SNR, as in Fig. \ref{fig: Figure 11}. The brown band indicates the $50\,\mathrm{MHz}-10\,\mathrm{GHz}$ range and the blue band denotes $0.1\,\mathrm{keV}-40\,\mathrm{keV}$.}
\label{fig: Figure 16}
\end{figure}
\textbf{SNR with $60\,M_\sun$ progenitor-} 
For self-consistent turbulence, the flux evolution of non-thermal emission in the different energy bands is similar to that for Bohm-like diffusion, which has been described in our previous study \myblue{\cite{2022A&A...661A.128D}}. Therefore, here we only summarise the difference in emission spectra and morphology on account of self-consistence turbulence and the time-dependent evaluation of diffusion coefficient.\\
Considerable synchrotron and gamma-ray emission emerge from the upstream region of the SNR forward shock already during the propagation of SNR in the shocked wind, on account of the escaped particles arising from the weak driving of turbulence, as indicated by Fig. \ref{fig: Figure 8}. The contact discontinuity between SNR forward and reverse shock appears bright in the radio and the X-ray band before the SNR approaches the shocked ISM, clearly visible in the X-ray morphology at 28000 years, illustrated in Fig. \ref{fig: Figure 18}. Around $28000$~years, following the collision of the SNR forward shock with the LBV shell, the wind-bubble contact discontinuity appears X-ray bright on account of the very strong magnetic field. The brightest radio emission still emerges from the region near the contact discontinuity between the SNR forward and reverse shock. The synchrotron morphology, illustrated in Fig. \ref{fig: Figure 18} modelled with self-consistent turbulence is similar to that for Bohm-like diffusion \myblue{\citep[cf.][]{2022A&A...661A.128D}}, except the contribution from escaped particles around the SNR.
This is to be noted that the radio spectra are softer with spectral index, $\alpha\approx 0.8$, where energy flux, $S_{\nu} = \nu^{-\alpha}$, during the shock passage through the shocked wind and the shocked ISM, which is quite consistent with the observed indices for many old Galactic SNRs \myblue{\citep{1977A&A....61...99B, 2009BASI...37...45G, 2014Ap&SS.354..541U}}.

\begin{figure}
\centering
\includegraphics[width=\columnwidth]{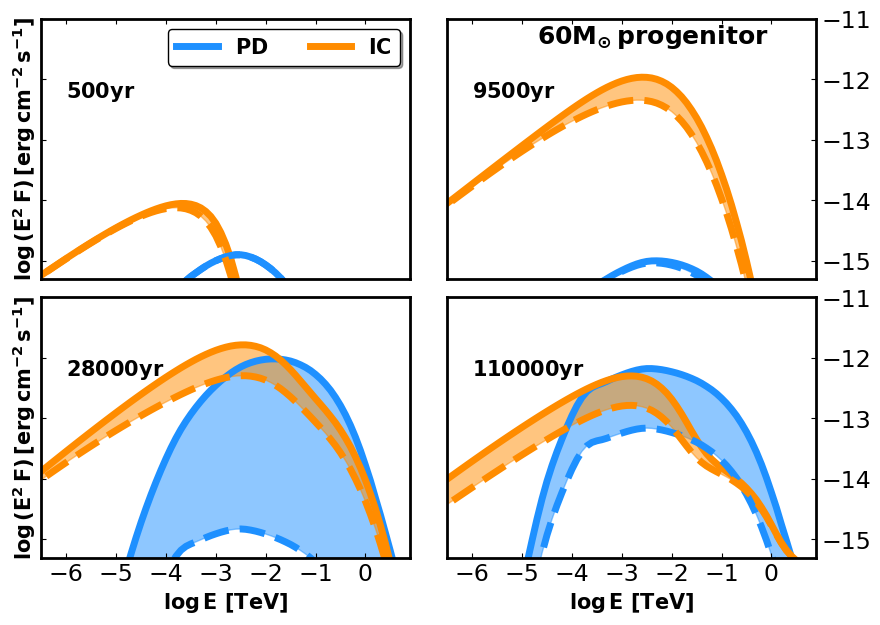}
\caption{\textbf{Gamma-ray emission by pion-decay (PD) and inverse Compton (IC) scattering at different ages.} The boundaries of the shaded regions indicate the internal and total emission, as in Fig. \ref{fig: Figure 11}.}

\label{fig: Figure 17}
\end{figure}
\begin{figure}[t]
\centering
\includegraphics[width=\columnwidth]{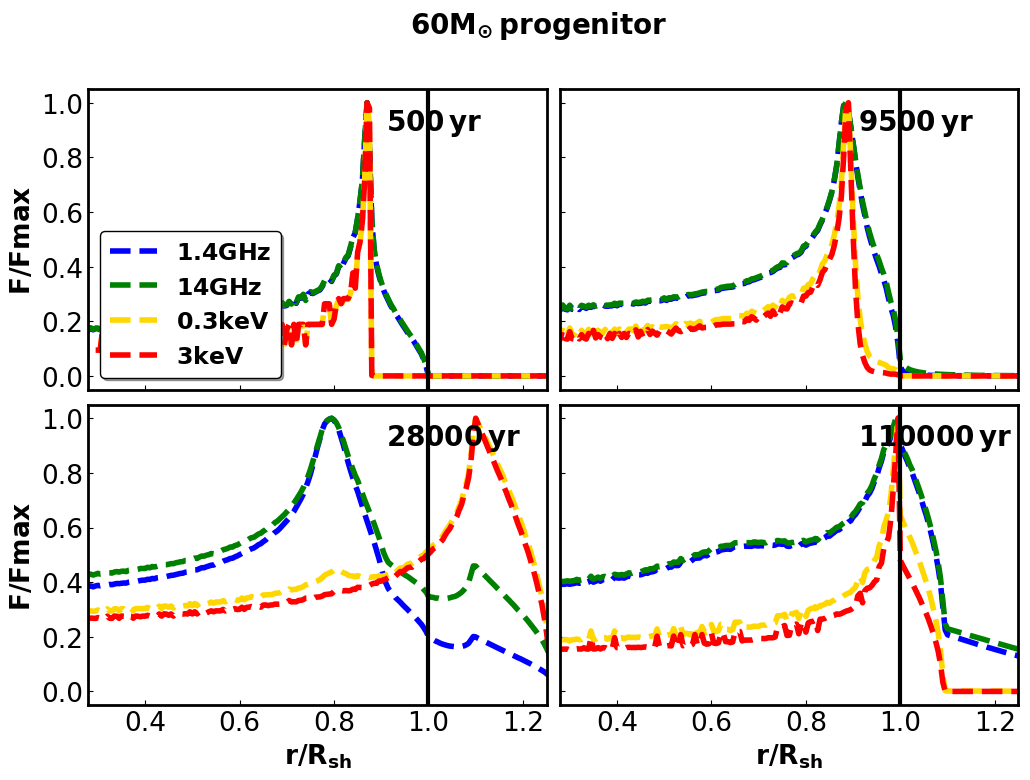}
\caption{\textbf{Normalised intensity profiles for synchrotron
emission.} For each specified profile, the intensity is normalised
to its peak value.}
\label{fig: Figure 18}
\end{figure}

\begin{figure}
\centering
\includegraphics[width=\columnwidth]{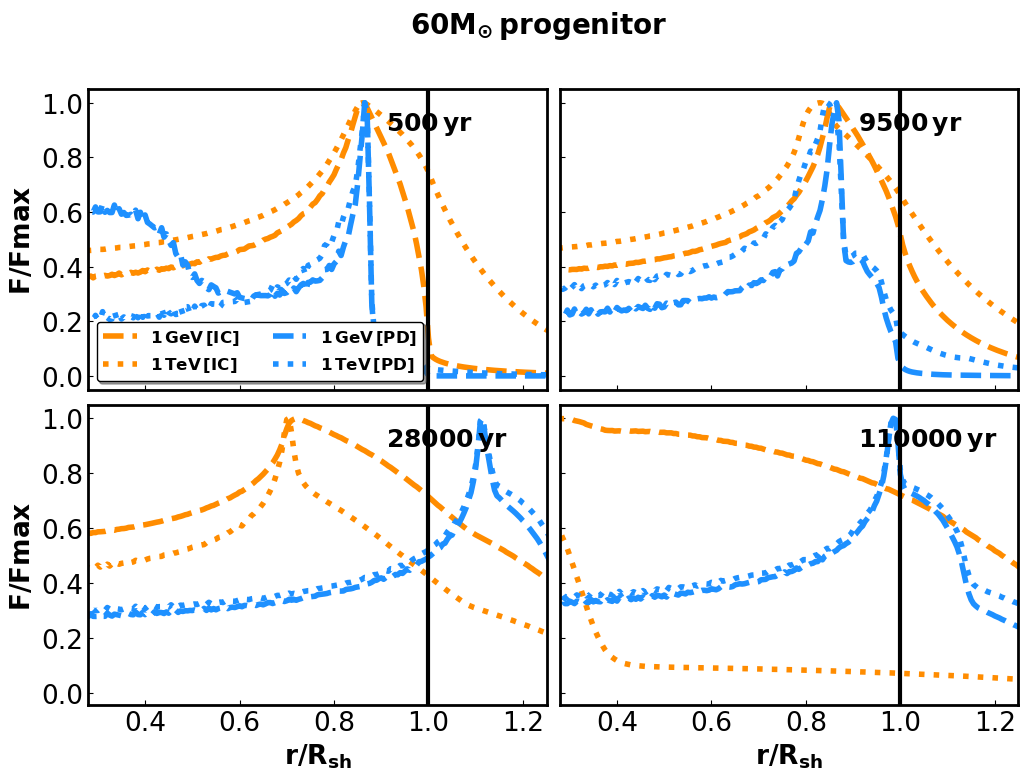}
\caption{\textbf{Normalised intensity profiles for gamma-ray
emission.} Emissions for pion-decay (PD) and inverse Compton (IC) emissions are shown. For each specified profile, the intensity is normalised to its peak value.}
\label{fig: Figure 19}
\end{figure}
During the early stages of evolution, the strong magnetic field in the free wind permits efficient proton acceleration to very high energy, and the gamma-ray emission is predominantly hadronic, independent of the diffusion model. At later stages, when the SNR expands in the shocked wind, inverse Compton emission dominates and pion-decay emission is diminished until particle escape becomes significant. Around $28000$~years, high-energy protons residing in the shock precursor can reach the high-density material behind the contact discontinuity of the wind bubble, which causes bright hadronic TeV-scale emission from the periphery of the SNR. Therefore, we obtain significant pion-decay emission flux from the upstream of the remnant, demonstrated in Fig. \ref{fig: Figure 17} at 28000 years and also visible in the morphology at $1\,\mathrm{TeV}$, shown in Fig. \ref{fig: Figure 19} where the pion-decay emission emerges from the shell around the wind-bubble contact discontinuity ahead of the SNR shock. This situation is comparable with the scenario of gamma-ray emission from the interaction of escaped protons with ambient molecular clouds, that has been suggested for IC 443, W44, $\mathrm{G\,39.2-0.3}$ and $\mathrm{G\,106.3+2.7}$ \myblue{\citep{10.1093/mnras/stt1318, 2022A&A...658A..60Y, 10.1093/mnras/staa2045}}. Pion-decay dominates the gamma-ray emission from old remnants whose forward shock propagates in the shocked ISM, and the spectral index ($ 2.2$-$2.4$) for pion-decay emission above 10 GeV reflects the softness of the proton spectra. 

The gamma-ray emission morphology is shell-like at early stages, dominated by the emission from the contact discontinuity between forward and reverse shock. After the collision with the wind bubble contact discontinuity, the velocity of SNR forward shock decreases, and it can no longer accelerate particles at very high energies. At that time, confined particles of very high energy can reach, and be re-accelerated by, the reverse shock that itself is energised by multiple reflected shocks resulting from the SNR-CSM interaction. It is already evident that after $28,000$ years, the inverse Compton emission mainly emerges from the region around the reverse shock, and the morphology eventually becomes centre-filled. This late-time morphology is similar to that obtained for the SNR with $20\,M_{\sun}$ progenitor.

\section{Conclusions}
We have probed the structures of the ambient medium at the pre-supernova stage for $20\,M_{\sun}$, and $60\,M_{\sun}$ progenitors by solving the hydrodynamic evolution of the CSM from the ZAMS stage to pre-supernova stage. We then have explored the interaction of the SNRs with the so-modelled CSM, as well as the acceleration and transport of energetic particles. For that purpose, we have solved the time-dependent transport equations of CRs in the test-particle limit and of magnetic turbulence, as well as the induction equation for the large-scale magnetic field, all in parallel with the hydrodynamic equations for the SNR and the CSM. The self-consistent turbulence module provides a time- and momentum-dependent spatial CR diffusion coefficient which is far more realistic than the oversimplified Bohm-scaling of the diffusion coefficient. 

We have demonstrated that inefficient confinement of high-energy particles eventually causes spectral breaks at GeV energies, above which the spectral index is $2.2-2.6$. The simulations of particle acceleration at the SNR forward shock indicate that the spectra of the particles and their emission products are significantly affected by the structure of the wind bubble. Our morphological analysis of two SNRs with $20\,M_{\sun}$, and $60\,M_{\sun}$ progenitors has revealed dissimilarities in various frequency bands, that  reflect the differences in the wind bubbles.

In the scenario with the $20\,M_{\sun}$ star, we have observed transient softer spectra with the spectral index of 2.2 specifically at higher energy during the collision between SNR-RSG shell. Beyond this, the flow profiles of the ambient medium do not induce any spectral softness. It is the weak driving of turbulence that for old remnants renders proton spectral indices around $2.6$ at high energies. Similar soft proton spectra have been deduced from observations of the SNRs evolving inside dense molecular clouds. The synchrotron flux depends on the total magnetic field intensity in the different regions of the wind bubble as well as the maximum achievable electron energy. At the later stage, although the magnetic field strength is very high in the shocked ISM, 
the paucity of high-energy electrons in the shock environment causes the synchrotron cut-off energy to decline from $10\,\mathrm{keV}$ to $0.1\,\mathrm{keV}$. The flux of pion-decay emission varies throughout the evolution, whereas the inverse Compton flux shows a steady trend and slightly increases until the SNR enters the shocked ISM. SNRs with a $20\,M_{\sun}$ progenitor have a shell-like morphology in the X-ray band and pion-decay emission except at middle age, while in the radio band the SNR appears more centre-filled, except at very old age. The inverse-Compton morphology is that of a thick shell, which transitions to a more centre-filled configuration at late times. 

For the SNR with $60\,M_{\sun}$ progenitor star, we have obtained soft spectra also at low energies, on account of the low sonic Mach number inside the hot shocked wind regions. At later times, the spectra become soft at higher energies as well, reflecting inefficient driving of turbulence and the associated rapid decline in the currently achievable maximum particle energy, that we already described for the SNR with $20\,M_{\sun}$ progenitor. The synchrotron cut-off frequency increases with time, on account of the high magnetic field intensity in the shocked wind, until the high-energy electrons start to escape from the remnant. Inverse Compton emission dominates the gamma-ray output until the SNR reaches the vicinity of the contact discontinuity of the wind bubble, and the pion-decay emission is prominent for the old remnant. For both types of SNR the gas density of the ambient medium determines the dominant contribution in the gamma-ray band \myblue{\citep[cf.][]{2012ApJ...761..133Y}}. The X-ray morphology resembles a thick shell, whereas the radio emission evolves with time from shell-like to centre-filled configuration. In the gamma-ray band, the pion-decay intensity profile is shell-like, and that of inverse Compton emission eventually transitions from shell-like to a centre-filled structure. For both SNRs, the morphology looks similar for old remnants, and in this stage, it also resembles that of type-Ia SNRs \myblue{\citep{2021A&A...654A.139B}}.

In conclusion, it is evident that the spectra of accelerated particles are shaped by both the hydrodynamics of the ambient medium and the time-dependent diffusion coefficients. Our results suggest that non-thermal emission and its morphology can inform us about the progenitor stars and the current state of evolution of the remnant, at least until it reaches the shocked ISM. The SNR with lower-mass progenitor star ($20\,M_{\sun}$) is more likely to be detected with current-generation observations, on account of the high density of the RSG wind. 

\begin{acknowledgements}
The authors acknowledge the North-German Supercomputing Alliance (HLRN) for providing HPC resources to perform the hydrodynamic simulation from ZAMS to pre-supernova stage of the star. Iurii Sushch acknowledges support by the National Research Foundation of South Africa (grant No. 132276). R. Brose acknowledges funding from the Irish Research Council under the Government of Ireland Postdoctoral Fellowship program.
\end{acknowledgements}

\bibliographystyle{aa}
\bibliography{ref}
\end{document}